\documentclass[onecolumn,showpacs,preprintnumbers,amsmath,amssymb,superscriptaddress]{revtex4-2}
\usepackage{graphicx}
\usepackage{dcolumn}
\usepackage{bm}
\usepackage{tabularx}
\usepackage{ulem} 
\newcolumntype{M}{>{\centering\arraybackslash}m{1.85cm}}
\usepackage[export]{adjustbox}
\usepackage{float}
\usepackage{color}   
\usepackage{xr-hyper}
\usepackage{hyperref}
\hypersetup{breaklinks=true,colorlinks=true,linkcolor=blue,citecolor=blue,filecolor=magenta,urlcolor=cyan}

\usepackage[all]{hypcap}

\usepackage{xcolor}
\definecolor{pastelgray}{rgb}{0.81, 0.81, 0.77}
\definecolor{beaublue}{rgb}{0.9, 0.9, 0.93}

\makeatletter
\def\@bibdataout@aps{%
	\immediate\write\@bibdataout{%
		@CONTROL{%
			apsrev41Control%
			\longbibliography@sw{%
				,author="08",editor="1",pages="1",title="0",year="1"%
			}{%
				,author="08",editor="1",pages="1",title="",year="1"%
			}%
		}%
	}%
	\if@filesw \immediate \write \@auxout {\string \citation {apsrev41Control}}\fi
}

\begin{document}
	
	\title{ \textit{Ab initio} no-core shell-model  study of $^{20-23}$Na isotopes}
	
	\author{ Chandan Sarma }
	\address{Department of Physics, Indian Institute of Technology Roorkee, Roorkee 247667, India}	
	\author{Praveen C. Srivastava}
	\email{Corresponding author: praveen.srivastava@ph.iitr.ac.in}
	\address{Department of Physics, Indian Institute of Technology Roorkee, Roorkee 247667, India}

	\date{\hfill \today}
	
	\begin{abstract}
		We have done a systematic no-core shell-model study of $^{20-23}$Na isotopes. The low-energy spectra of these sodium isotopes consisting of natural and un-natural parity states were reported, considering three realistic interactions: inside nonlocal outside Yukawa (INOY), charge-dependent Bonn 2000 (CDB2K), and the chiral next-to-next-to-next-to-leading order (N$^3$LO). We also present the mirror energy differences in the low-energy spectra of $|T_z|$ = 1/2 mirror pair ($^{21}$Na - $^{21}$Ne). Apart from the energy spectra, we have also reported the electromagnetic transition strengths and moments. Finally, considering all three realistic interactions, we report the point-proton radii and neutron skin thicknesses.
	\end{abstract}
	
	
	\maketitle
	
	\section{Introduction}
	\label{sect 1}
	For the last two decades, the no-core shell-model (NCSM) \cite{NCSM_r2,NCSM_r1,MVS2009,A=10,C.Forssen,stetcu1,stetcu2,Dytrych,McCoy} became a major \textit{ab initio} approach to solve nuclear many-body problems starting from realistic $NN$ and $3N$ interactions. The NCSM approach is very successful in explaining nuclear structural properties of $p$-shell nuclei \cite{Phys.Rev.C573119(1998),Phys.Rev.C502841(1994), PRL84, PRC62,  Choudhary1,Choudhary2,Choudhary3,arch1,Phys.Rev.C54001(2021),reviewJPG,NCSM_p8}, although it is less explored for the $sd$-shell nuclei mainly due to the significant increase in computational resources.  Some approximations become necessary, 
	in terms of small model spaces considered, to tackle lower $sd$-shell nuclei within the NCSM formalism. Like any other many-body methods, employed approximations would cause deficiencies in reproducing the correct results of some of the nuclear observables. However, understanding the nature of such deficiencies is important for improving the nuclear Hamiltonians taken into account to study nuclei of a particular region of the nuclear chart.

{\color{black} The knowledge of low-energy states, along with electromagnetic properties, provides insight into the internal structures of atomic nuclei. Additionally, nuclei near $ N = Z $ serve as ideal test cases for various many-body approaches in nuclear structure studies.}    
{\color{black}Previously, we investigated the neon isotopic chain from $ A = 18 $ to $ A = 24 $ using the NCSM formalism \cite{chandan}. The computed energy spectra, $ B(M1) $, and magnetic moments show good agreement with experimental data. However, observables that depend on the long-range components of nuclear wavefunctions, such as $ B(E2) $ and the quadrupole moment, exhibit significant deviations from experimental results.
To check the consistency of the NCSM formalism in describing lower $sd$-shell nuclei, we now focus on the sodium isotopic chain for which several theoretical as well as experimental investigations were carried out in recent times. }
	Recently, the $ E2$ {\color{black}transition} strengths between the ground and the first excited state of $^{23}$Na and its mirror nucleus $^{23}$Mg are measured using Coulomb excitation measurement \cite{vs_imsrg1}. The experimental $E2$ strengths of both nuclei are compared to the $ab$ $initio$ valence space in-medium similarity group (VS-IMSRG) results and standard shell model results within the $ sd$ space. A similar kind of theoretical investigation is reported in Ref. \cite{vs_imsrg2} for a range of mirror nuclei pairs across the $sd$-space, including $|T_z|$ = 1/2 mirror pairs ($^{21}$Na-$^{21}$Ne) and ($^{23}$Na-$^{23}$Mg). This study shows that missing $E2$ strengths for VS-IMSRG are mostly isoscalar in nature. In Ref. \cite{arch3}, the electric quadrupole and magnetic dipole moments of $^{20-31}$Na isotopes are studied along with other nuclei across the $sd$-space using valence-space Hamiltonian constructed from {\it ab initio} VS-IMSRG and coupled cluster effective interaction (CCEI).
	Recently, the isospin-symmetry breaking in the mirror-energy difference (MED) of $sd$- and $pf$-shell nuclei are investigated in Ref. \cite{MED} within the {\it ab initio } VS-IMSRG. This study shows that while the calculated MEDs of $^{20}$Na ($1^+_2$) and $^{21}$Na (1/2$^+_1$) are close to the experimental MEDs, the calculated MED of $^{19}$Na (1/2$^+_1$) is small compared to the experimental one. On the other hand, the charge radius is another fundamental property of atomic nuclei, apart from low-energy spectra and electromagnetic observables, that can reveal the structure of nuclei to a great extent.  In recent years, several experiments have been performed to measure the charge radii of different isotopic chains. On the other hand, the calculated charge radii with different many-body theories having error quantification procedures facilitate the comparison of theoretical and experimental results directly. Recently, the charge radii of Na isotopes have been extracted from accurate atomic calculations in Ref. \cite{Na_rc}. Suzuki \textit{et. al. } in Ref. \cite{Na_skin} studied the development of neutron skin along the Na isotopic chain. Using shell-model, the variation of charge and matter radius along the sodium isotopic chain is reported in Ref. \cite{subhrajit}. 

	In this work, we have performed nuclear structure study of $^{20-23}$Na isotopes within the formalism of \textit{ab initio} no-core shell model using three realistic interactions, namely inside nonlocal outside Yukawa (INOY) \cite{INOY, INOY2, INOY3}, charge-dependent Bonn 2000 (CDB2K) \cite{CDB2K1, CDB2K2}, and the chiral next-to-next-to-next-to-leading order (N$^3$LO) \cite{QCD, N3LO}. We have calculated the  ground state (g.s.) energies, low-energy spectra comprising of both natural and un-natural parity states, electromagnetic transition strengths and moments, and also the point-proton radii of these sodium isotopes. The maximum basis space we have reached in this work is $N_{max}$ = 4 for all four sodium isotopes. This work is important to check the consistency of our previous works on the neon chain \cite{chandan} as well as other works on the lower mass $sd$-shell nuclei \cite{arch1, arch2} using the NCSM formalism. 
	
	Present  work is organized as follows: In \autoref{sect 2}, the basic formalism of the NCSM approach is given. The $NN$ interactions used in this work are briefly described in \autoref{sect 3}. In \autoref{sect 4}, the NCSM results for energy spectra, electromagnetic properties, and point-proton radii are presented. Finally, we summarize our work in \autoref{sect 5}.
	
	\section{Formalism of no-core shell-model}
	\label{sect 2}
	
	In this approach, nucleons are considered to be non-relativistic point particles interacting via realistic $NN$
	and $NN$ + $3N$ interactions. In the present work, we are using only the $NN$ interactions for which the Hamiltonian can be written as follows \cite{NCSM_r2}:
	\begin{eqnarray}
	\label{eq:(1)}
	H_{A}= T_{rel} + V = \frac{1}{A} \sum_{i< j}^{A} \frac{({\vec p_i - \vec p_j})^2}{2m}
	+  \sum_{i<j}^A V^{NN}_{ij}, 
	\end{eqnarray}
	where, $T_{rel}$ is the relative kinetic energy and $V^{NN}_{ij}$ is the $NN$ interaction containing nuclear part as well as Coulomb part.
	
	In the NCSM approach, the solution of a non-relativistic Schr\"odinger equation for an A-nucleon system is found by performing a large-scale matrix diagonalization in a many-body Harmonic oscillator basis. The many-body basis is constructed by taking the Slater determinant (M-scheme) of the single-particle Harmonic oscillator orbitals having a truncation parameter $N_{max}$. 
    {\color{black}The $N_{max}$ shows the number of allowed harmonic-oscillator quanta taken above the minimum configuration allowed by the Pauli exclusion principle. }
     The use of the harmonic oscillator many-body basis, along with $N_{max}$-truncation, allows the smooth separation of the center of mass (CM) coordinates from the relative coordinates. Large $N_{max}$ calculations are necessary to obtain converged results while dealing with hard-core potentials having strong short-range correlations. However, large $N_{max}$ calculations are computationally challenging, thus we need some renormalization technique to soften such potentials.  Two such renormalization procedures that are popularly used in NCSM calculations are the Okubu-Lee-Suzuki (OLS) \cite{OLS1, OLS2, OLS3, OLS4} transformation and similarity renormalization group (SRG) \cite{SRG1, SRG2}. Both these procedures soften the hard-core part of nuclear interactions eventually to obtain converged results within a computationally tractable model space controlled by the $N_{max}$ parameter. In this work, we are using the OLS technique to obtain an effective Hamiltonian. 
	
	To facilitate the derivation of the OLS effective Hamiltonian, the center-of-mass (c.m.) Hamiltonian, $H_{c.m.}$ is added to the original Hamiltonian \autoref{eq:(1)}: 
	
	\begin{eqnarray}
	\label{eq:(2)}
	H^{\Omega}_A = H_A + H_{c.m.} = \sum_{i = 1}^A [\frac{p_i^2}{2m} + \frac{1}{2}m \Omega^2 r_i^2]
	+ \sum_{i < j}^A [V_{NN, ij} - \frac{m \Omega^2}{2 A} (r_i - r_j)^2].
	\end{eqnarray}
	
	Here, $H_{c.m.}$ = $T_{c.m.}$ + $U_{c.m.}$ and $T_{c.m.}$ and $U_{c.m.}$ are the kinetic and potential terms for the center of mass co-ordinate. $H_A$, being translationally invariant which does not change the intrinsic properties once $H_{c.m.}$ is added to it.
	
	In order to develop effective interactions, first the infinite HO basis is separated into $P$- and $Q (=1-P) $ spaces by using projection operators.
	The $P$-space contains all the HO orbitals allowed by the truncation parameter $N_{max}$ and the $Q$-space contains all the excluded HO orbitals. The effective Hamiltonian is then constructed by using the OLS unitary transformation on $H^{\Omega}_A$ given in \autoref{eq:(2)}. The OLS transformation induces up to A-body terms even if we start with the $NN$ Hamiltonian given in \autoref{eq:(1)}. In this work, we are considering only up to 2-body terms as contributions from those terms would be dominant over other many-body terms.  In the final step, the $H_{c.m.}$ is subtracted, and the Lawson projection term, $\beta (H_{\mbox{c.m.}} - \frac{3}{2}\hbar\Omega)$ \cite{Lawson} is added to the Hamiltonian. In this work, the value of $\beta$ is taken to be 10 and the addition of the Lawson projection terms shifts energy levels arising due to the excitation of the CM. So, the final form the Hamiltonian can be written as: 
	
	\begin{multline}
	\label{eq:(3)}
	H_{A,eff}^{\Omega} = P\left\{ \sum _{i<j}^{A} \left[ \frac{{(\vec p_i - \vec p_j)}^2}{2mA} + \frac{m {\Omega}^2}{2A} {(\vec r_i - \vec r_j)}^2 \right]\right. 
	\left.+\sum_{i<j}^{A} \left[ V^{NN}_{ij} - \frac{m {\Omega}^2}{2A}{(\vec r_i - \vec r_j)}^2\right]_{\rm eff} + \beta \Bigg(H_{\mbox{c.m.}} - \frac{3}{2}\hbar\Omega\Bigg)\right\}P.\\
	\quad
	\end{multline}
	
	The effective Hamiltonian of \autoref{eq:(3)} is dependent on the number of nucleon (A), HO frequency ($\Omega$), and number of HO orbitals considered in the $P$-space controlled by the truncation parameter, $N_{max}$. In order to minimize the effect of many-body terms in developing an effective Hamiltonian, a large model space is needed for NCSM. For the extreme limit of $N_{max} \rightarrow \infty$, the effective Hamiltonian of \autoref{eq:(3)} reduces to the bare Hamiltonian of \autoref{eq:(1)} and the NCSM results with effective Hamiltonian reduce to the exact solution.

	\section{Realistic $NN$ interactions}
	\label{sect 3}
	In this work, we have performed NCSM calculations using three realistic $NN$ interactions namely, inside nonlocal outside Yukawa (INOY) \cite{INOY, INOY2, INOY3}, charge dependent Bonn $NN$ interaction (CDB2K) \cite{CDB2K1, CDB2K2}, and the chiral N$^3$LO \cite{QCD, N3LO}. 
	
	The inside nonlocal outside Yukawa (INOY) \cite{INOY, INOY2, INOY3} is one of the interactions we use in our current study. It is a phenomenological $NN$ potential that can simultaneously reproduce the $NN$ data and $3N$ bound states without an additional $3N$ force. The Yukawa tail of the local Argonne $v18$ \cite{av18} potential is smoothly cut off in 1-3 fm region, and a non-local potential is added and later on fitted to experimental data. The full INOY interaction is defined as \\
	
	V$^{full}_{ll'}(r,r')$  = W$_{ll'}(r,r')$ + $\delta (r-r')$ F$^{cut}_{ll'}(r)$ V$^{Yukawa}_{ll'}(r)$, where $W_{ll^{'}}(r,r^{'})$ and $V_{ll^{'}}^{Yukawa}(r)$ are, respectively, the nonlocal part and the local part of this interaction. The cut-off function F$^{cut}_{ll'}(r)$ is defined as
	
	\[ 
	F^{cut}_{ll'}(r) =
	\begin{cases}
	1- e^{-[\alpha_{ll'}(r-R_{ll'})]^{2}} & for ~r\geq R_{ll'} ,\\
	0 & for ~r\leq R_{ll'}. \\
	\end{cases}
	\]
	
{\color{black}The parameters $\alpha_{ll'}$ and $R_{ll'}$ are considered to be independent of the angular momenta having values 1.0 $fm^{-1}$ and 1.0 $fm$, respectively. Additionally, different parameters of the non-local part, W$_{ll'}(r,r')$, are determined by fitting $NN$ data and the binding energy of $^3$He}. The INOY interaction includes charge independence breaking (CSB) and charge symmetry breaking (CIB) as it is fitted to reproduce all low-energy experimental parameters, including $np$, $pp$, and $nn$ scattering lengths, up to high precision. The advantage of using INOY interaction over the other local $NN$ interactions is that it does not require an additional $3N$ force to provide the binding energies of $3N$ systems. The non-local part of the interaction, $W_{ll^{'}}(r,r^{'})$, partly absorbs the contributions from $3N$ force, and the internal structure of nucleons mainly contributes to the non-locality part of this $NN$ interaction.
	
	The second interaction employed in this work is the charge-dependent Bonn 2000 (CDB2K) potential that is fitted to the world $pp$ data available in the year 2000. This is a one-boson-exchange (OBE) $NN$ interaction that takes into account all mesons ($\pi$, $\eta$, $\rho$, $\omega$) with masses less than the nucleon mass.  However, the $\eta$-meson is dropped from the potential as it has a vanishing coupling with the nucleon. Additionally, two scalar-isoscalar $\sigma$ bosons ($\sigma_1$ and $\sigma_2$) are also included in the potential model. Both $\sigma_1$ and $\sigma_2$ are partial wave dependent and describe the intermediate range attraction among nucleons. The CDB2K interaction includes CSB and CIB in all partial waves up to $J = 4$ as the Bonn full model \cite{CDB2K2}. The interaction contains three parts: proton-proton ($pp$), neutron-neutron ($nn$), and neutron-proton ($np$), and the $T = 1$ parts of these potentials are related by CSB and CIB.  The $T = 0$ part of the $np$ potential is fitted separately. The original form of the covariant Feynman amplitudes without local approximation is used in the CDB2K interaction for meson exchange, and it provides a different off-shell behavior for this $NN$ potential compared to other local $NN$ potentials.

	The third interaction used in our current study is a chiral $NN$ interaction at next-to-next-to-next-to-leading order (N$^3$LO) \cite{N3LO} derived from the quantum chromodynamics (QCD) using chiral perturbation theory ($\chi$PT). The accuracy of this interaction in reproducing the $NN$ data below 290 MeV lab energy can be compared to the results of high-precision phenomenological potential AV18 \cite{av18}. In $\chi$PT, a systematic expansion of the nuclear potential is done in terms of (Q/$\Lambda_\chi$)$^\nu$, where Q and $\Lambda_\chi$ are a low-momentum scale or pion mass and the chiral symmetry breaking scale ($\approx$ 1 GeV), respectively. For a definite order $\nu$ ($\geq$ 0), there are a finite number of contributing terms to the nuclear potential. The exchange of pions contributes to the long-range part of nuclear interactions, while the contact terms are associated with the short-range part. Also, a certain number of charge-dependent parameters are also included in nuclear potential.  There are, in total, 29 parameters of the N$^3$LO potential: three of them are low-energy constants (LECs), $c_2$, $c_3$, and $c_4$ related to the $\pi N$ Lagrangians, 24 contact terms that dominate the partial waves with L $\le$ 2, and the remaining two parameters are two charge-dependent contact terms. 
	
	 In the present work, NCSM calculations have been performed with the pAntoine code \cite{pAntoine1,pAntoine2,pAntoine3}.

	\section{Results and Discussions}
	\label{sect 4}
	In this section, we have reported NCSM results of different nuclear observables. The NCSM calculations are computationally challenging for medium mass nuclei mainly due to the huge dimensions of the Hamiltonian matrix. In \autoref{tab:dimension}, the $M$-dimensions of the Hamiltonian matrices are shown for $^{20-23}$Na isotopes corresponding to the model space from $N_{max}$ = 0 to 6. We are able to reach basis size up to $N_{max}$ = 4 for all four sodium isotopes (shown in blue color in \autoref{tab:dimension})  with the available computational resources. The maximum dimension we are able to reach is 1.1 $\times$ 10$^9$ for $^{23}$Na corresponding to $N_{max}$ = 4 model space.
	
\begin{figure*}
		\centering
		\includegraphics[scale = 0.57]{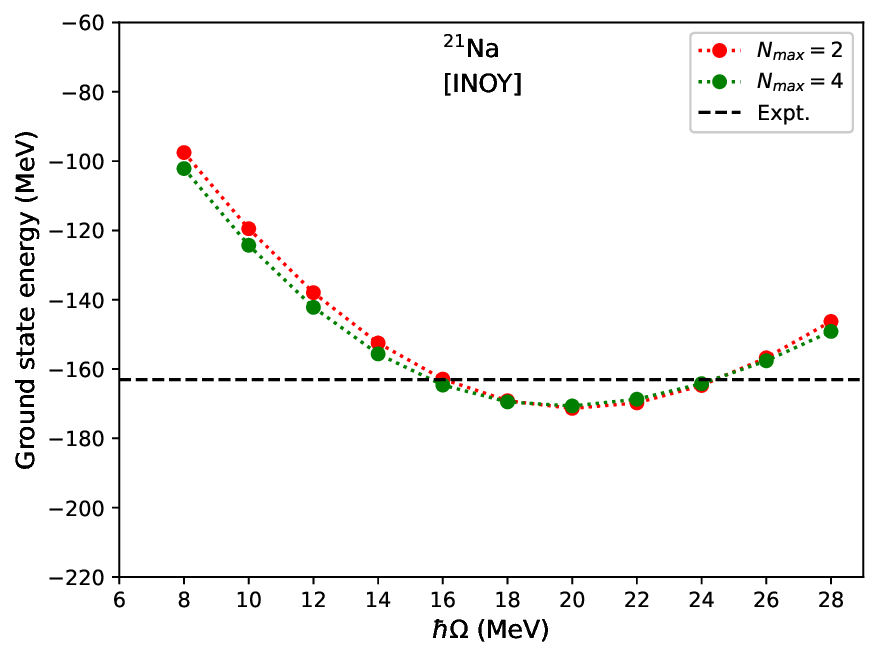}
		\hspace{0.75cm}
		\includegraphics[scale = 0.57]{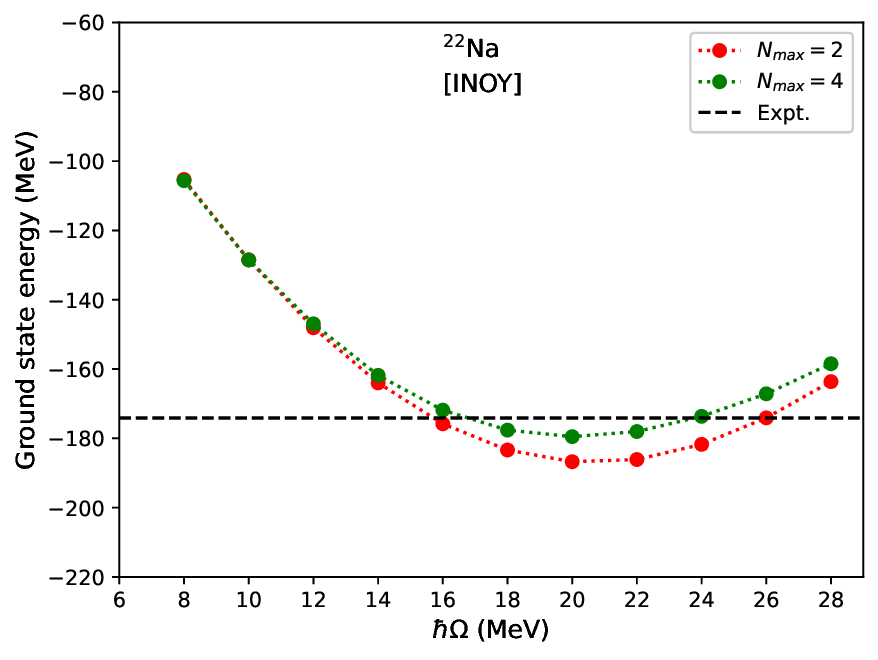}
		\includegraphics[scale = 0.57]{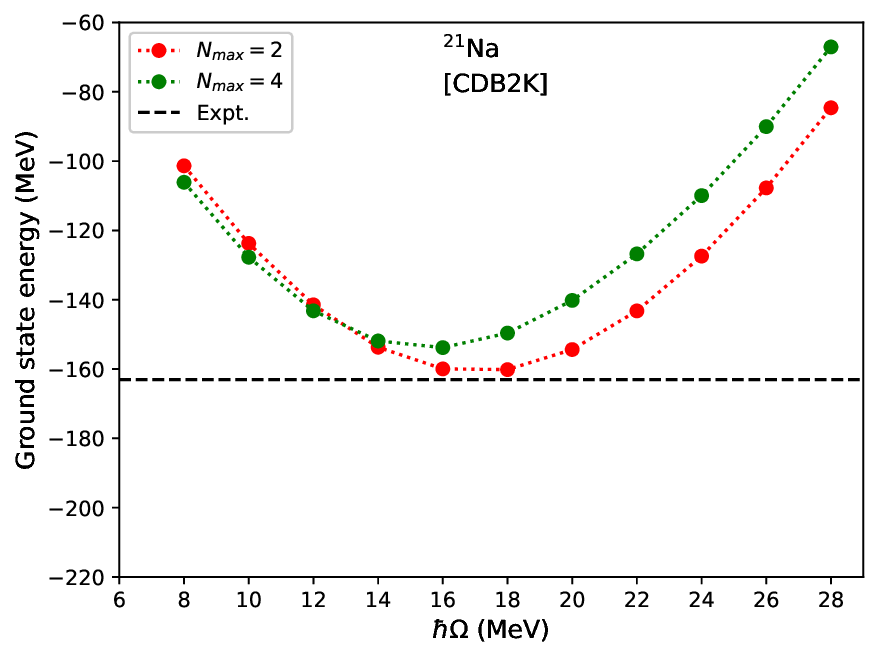}
		\hspace{0.75cm}
		\includegraphics[scale = 0.57]{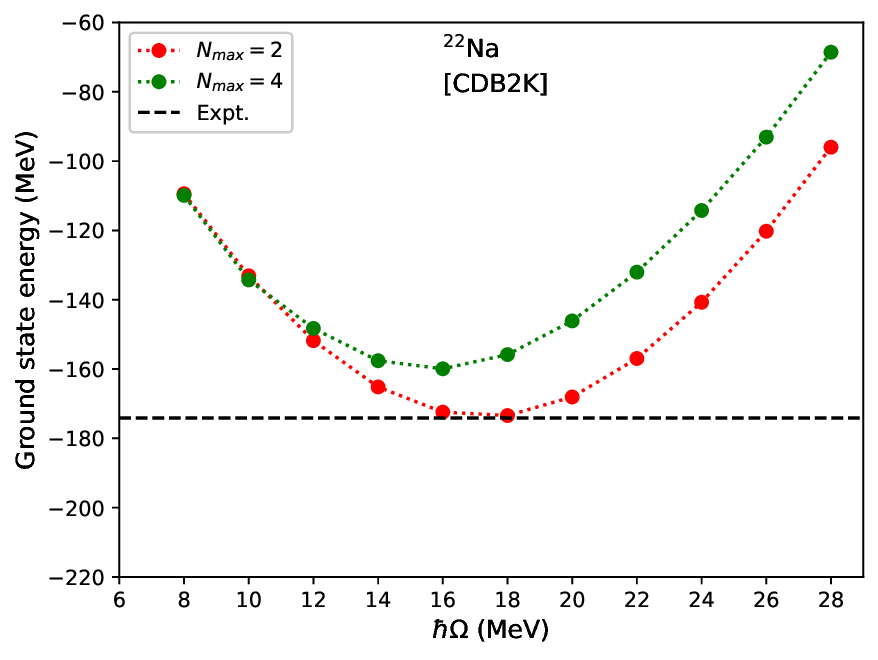}
		\includegraphics[scale = 0.57]{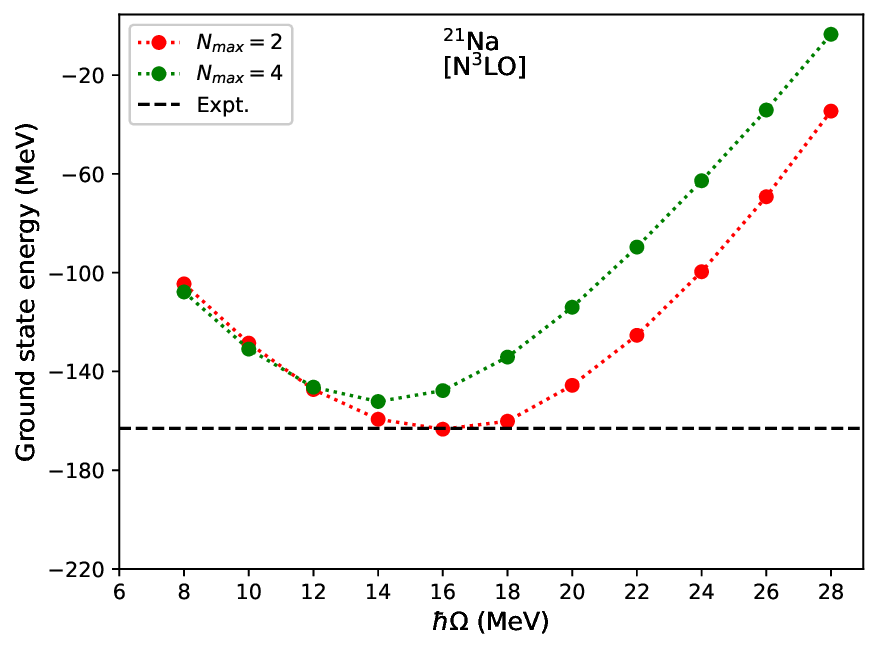}
		\hspace{0.75cm}
		\includegraphics[scale = 0.57]{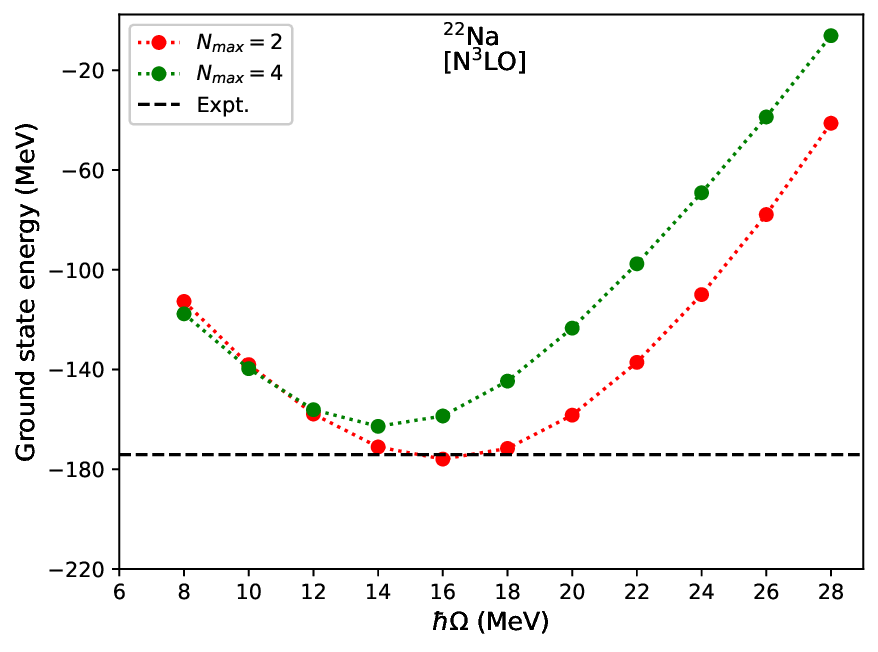}
		\caption{The variation of calculated g.s. energies of $^{21,22}$Na with HO frequencies for different $N_{max}$ are shown corresponding to three different $NN$ interactions. The horizontal lines show the experimental binding energies of the g.s.}
		\label{gs}
	\end{figure*}

\begin{figure*}
		\centering
		\includegraphics[scale = 0.57]{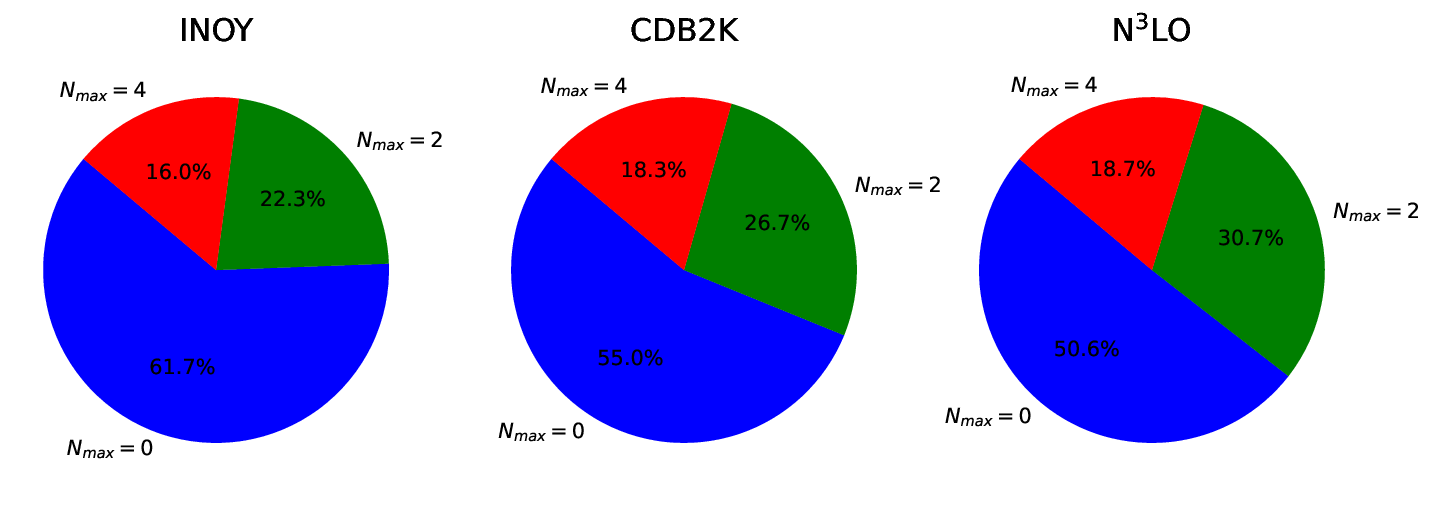}
		\caption{Probability distributions over 0$\hbar \Omega$ (in blue), 2$\hbar \Omega$ (in green), and 4$\hbar \Omega$ (in red) subspaces are shown for the $N_{max}$ = 4 NCSM g.s.  of $^{23}$Na considering three interactions.}
		\label{Pi_occupancy}
	\end{figure*}

	The first step of the NCSM calculation is to decide an optimum frequency for which different observables are calculated later on. 
	The optimum frequency for a particular interaction is decided by plotting the ground state energies for a particular interaction corresponding to different $N_{max}$.
	Subsequently, the frequency corresponding to the minimum g.s. energy calculated at the highest $N_{max}$ is taken as the optimum frequency for a particular interaction. 
	In \autoref{gs}, the g.s. energies of $^{21, 22}$Na isotopes are shown with realistic $NN$ interactions: INOY, CDB2K and N$^3$LO for $N_{max}$ = 2 and 4 model spaces. The optimum frequencies for INOY, CDB2K, and N$^3$LO for both isotopes are 20, 16, and 14 MeV, respectively. Similarly, the optimum frequencies for other Na-isotopes are calculated. {\color{black} From \autoref{gs}, it is evident that the g.s.  binding energies do not exhibit variational behavior concerning the truncation parameter, $N_{\text{max}}$, and the harmonic oscillator frequency, $\hbar \Omega$. While the NCSM results with bare interactions are variational, converging from above with $N_{max}$ considering $\hbar \Omega$ as the variational parameter \cite{NCSM_r2}, the results for OLS transformed interactions are not variational as some part of the bare interactions is omitted.}
    In \autoref{Pi_occupancy}, we have shown the probability distributions over different model spaces from $N_{max}$ = 0 to 4 for the NCSM calculation of $^{23}$Na g.s. for $N_{max}$ = 4 calculation. From the figure, we see that a maximum contribution of 18.7\% comes from $N_{max}$ = 4 space for the N$^3$LO interaction among the three interactions. 
	
	\begin{table}[ht]
		\centering
		\caption{ The dimensions of the g.s.  of Na isotopes corresponding to different $N_{max}$ are shown. The dimensions up to which we have reached are shown in blue.}
		\label{tab:dimension}
		\begin{tabular}{ccccc}
			\hline
			\hline\\
			$N_{max}$  &  $^{20}$Na    &  $^{21}$Na    &  $^{22}$Na   & $^{23}$Na \\
			\hline \vspace{-2.9mm}\\
			0 & {\color{blue} 394}   & { \color{blue} 1935} & {\color{blue} 6116} & {\color{blue} 1.3 $\times 10^4$}\\
			2 & {\color{blue} 4.0 $\times 10^5$} & {\color{blue} 1.5 $\times 10^6$} & {\color{blue} 4.6 $\times 10^5$} & {\color{blue} 8.4 $\times 10^6$}\\
			4 & {\color{blue} 5.9 $\times 10^7$} & {\color{blue} 1.9 $\times 10^8$} & {\color{blue} 5.2 $\times 10^8$} & {\color{blue} 1.1 $\times 10^9$}\\
			6 & {\color{black} 3.6 $\times 10^9$} & {\color{black} 1.1 $\times 10^{10}$} & {\color{black} 3.0 $\times 10^{10}$} & {\color{black} 6.5 $\times 10^{10}$}\\
			\hline \hline
		\end{tabular} 
	\end{table}

	After deciding the optimum frequency for a particular interaction, the low-energy spectra and other nuclear observables are calculated corresponding to the optimal frequency. In order to compare the results of different interactions for the low-energy spectra of a particular nucleus, we use root mean square deviation (rms) of energy spectra defined as:
	\begin{eqnarray}
	\label{eq:(4)}
	E_{rms} = \sqrt{\frac{1}{N}\sum_{i}^{N}(E_{exp}^i - E_{th}^i)^2}.
	\end{eqnarray}
	
	Here, $E_{exp}^i$ and $E_{th}^i$ are, respectively, the experimental and the calculated energy for the $i^{th}$ state of a particular nucleus. The $N$ is the number of states considered for calculating the rms value, and the $E_{rms}$ provides the quality of theoretical results compared to the experimental data. 
	
	\subsection{Natural-parity low-lying states of $^{20-23}$Na:}
	
	\begin{figure*}
		\includegraphics[scale = 0.55]{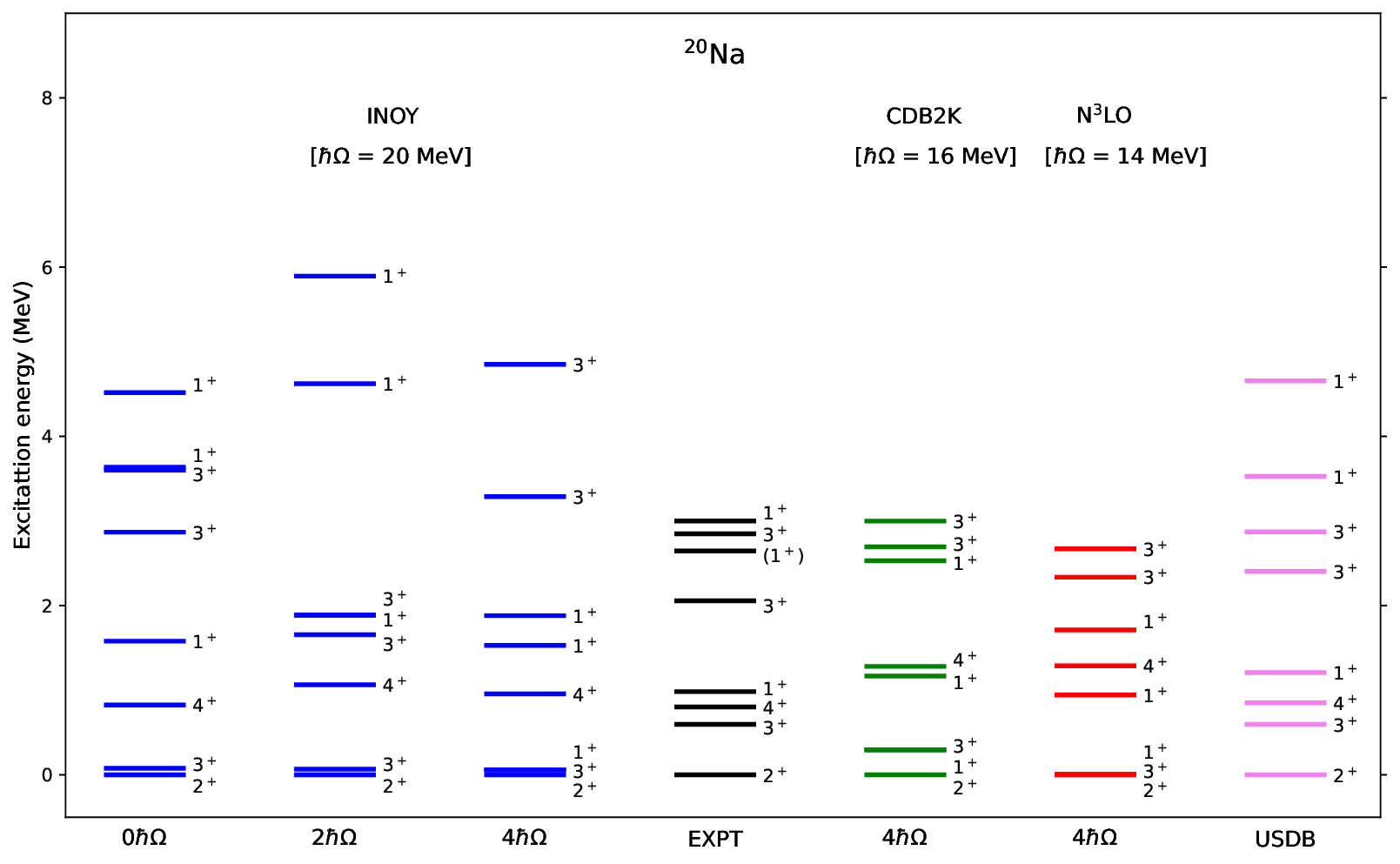}
		\includegraphics[scale = 0.55]{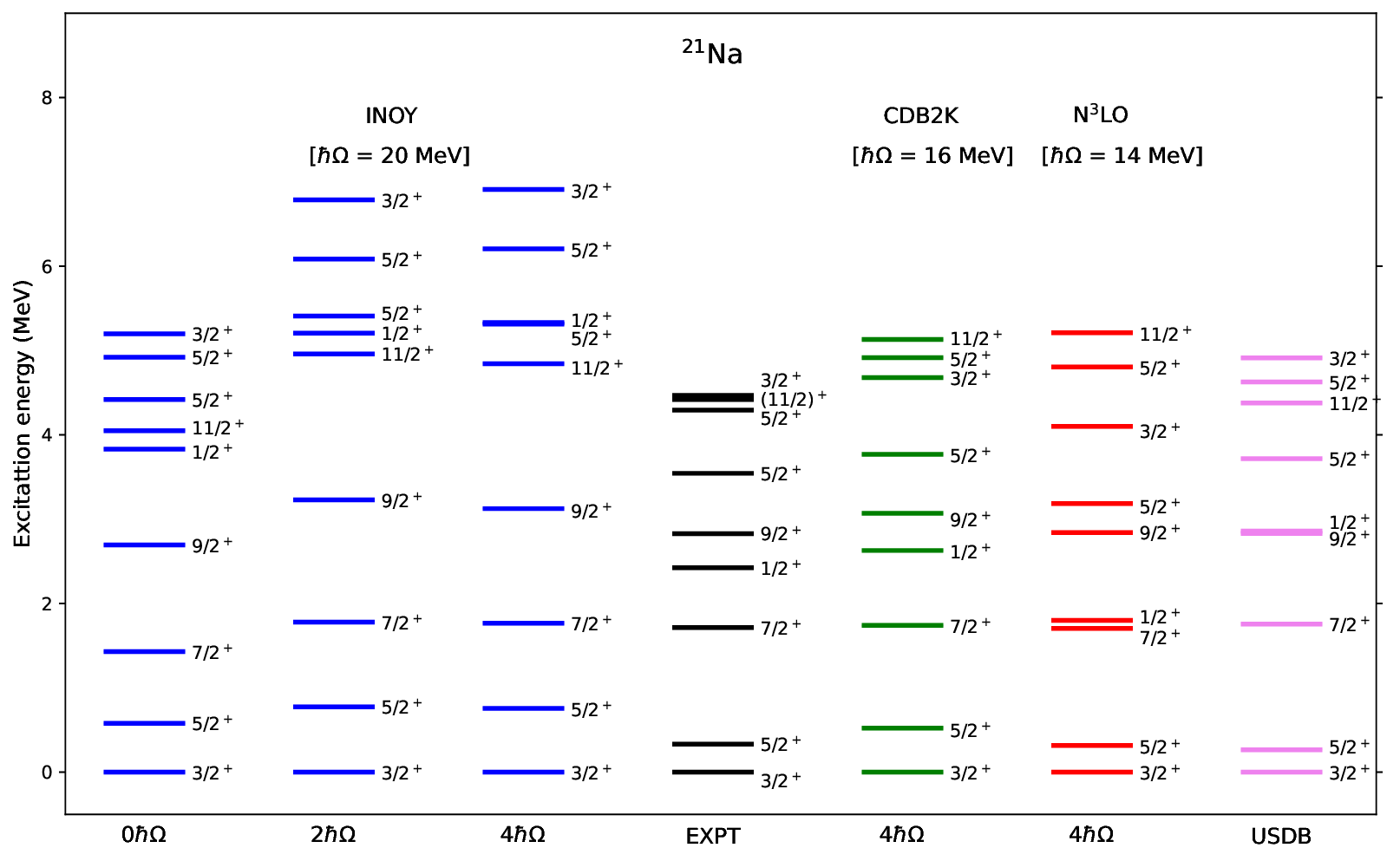}
		\caption{Natural parity low-energy spectra of $^{20, 21}$Na.}
		\label{spectra1}
	\end{figure*}
	
	\begin{figure*}
		\centering
        \includegraphics[scale = 0.55]{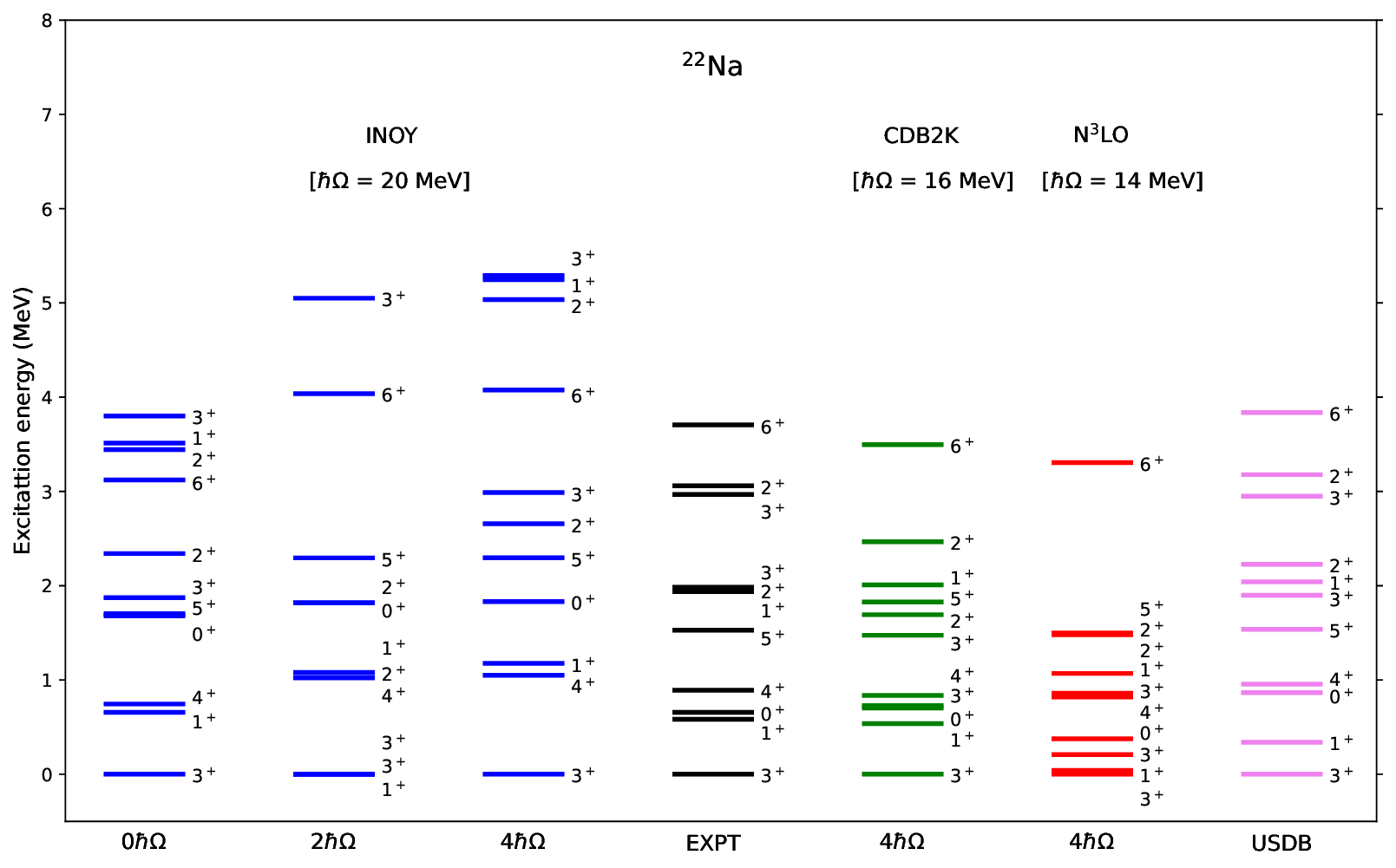}
		\includegraphics[scale = 0.55]{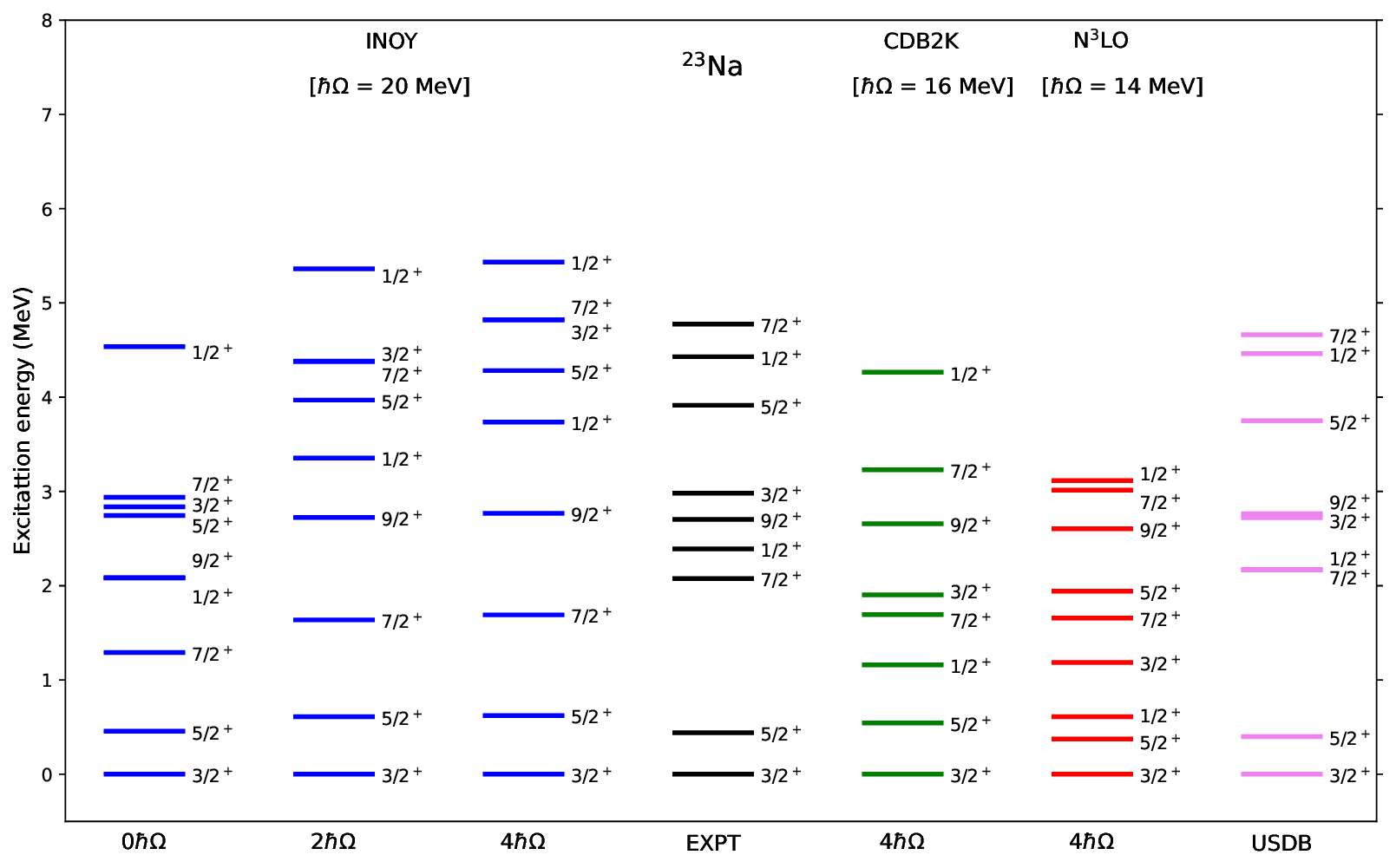}
		\caption{Natural parity low-energy spectra of $^{22, 23}$Na.}
		\label{spectra}
	\end{figure*}

	The first panel of \autoref{spectra1} shows the positive-parity low-lying spectra of $^{20}$Na. As shown in the figure, 2$^+$ is the experimental ground state, which is well reproduced by all three interactions, namely INOY, CDB2K, and N$^3$LO. However, the calculated  g.s. energies are only a few keV lower than the energies of the first excited states for all three interactions. The INOY and N$^3$LO interactions correctly reproduce the first excited state of $^{20}$Na though the excitation energies are 58- and 7- keV, respectively, compared to the experimental excitation energy of 596 keV. None of the three realistic interactions are able to reproduce the excitation spectra above the first excited state. 
	However, the excitation energies of $4_1^+$ state obtained for INOY and $3_3^+$ state obtained for CDB2K are only 150 keV away from the experimental data.
	The rms deviation in energies for the eight lowest states are 1.146, 0.720, and 0.941 MeV, for INOY, CDB2K, and N$^3$LO $NN$ interactions, respectively. {\color{black} We also compared the shell-model results using well-known USDB \cite{usdb} interaction to the NCSM results. The shell-model results show that while the ordering of 2$^+_1$-3$^+_1$-4$^+_1$-1$^+_1$ states is well reproduced, other four states are seen at higher energies compared to the experimental data. }
	
	The low-lying spectra of $^{21}$Na for INOY, CDB2K, and N$^3$LO interactions are shown along with the experimental data in the second panel of \autoref{spectra1}. The figure shows that the CDB2K and N$^3$LO reproduce the correct ordering of low energy states up to the fifth excited state ($5/2^+$). On the other hand, the INOY interaction is able to reproduce correct ordering only up to the second excited state (7/2$^+$). 
	A significant deviation of 2.907 MeV is observed between the calculated excitation energy with INOY and the experimental data for the $1/2_1^+$  state.
	The root-mean-square deviations corresponding to nine low-lying states between the experimental and calculated results for the INOY, CDB2K, and N$^3$LO are 1.645, 0.375, and 0.439 MeV. However, the rms deviations are 0.336, 0.388, and 0.396 MeV corresponding to INOY, CDB2K, and N$^3$LO for the yrast band states (3/2$^+$-5/2$^+$-7/2$^+$-9/2$^+$-11/2$^+$). On the other hand, the shell-model results using USDB interaction show good agreement with the experimental data.
	
	For the case, $^{22}$Na, the g.s.  (3$^+$) is well reproduced by all the three interactions considered in this work as shown in the first panel of \autoref{spectra}. While the CDB2K interaction is able to reproduce low-energy spectra up to the second excited state (0$^+$), the N$^3$LO interaction is able to reproduce only the g.s.  and the first excited state in the correct order. The calculated excitation energies of the 4$^+$ state with CDB2K and N$^3$LO interactions are only 56 and 73 keV away from the experimental data. The rms deviations in energies for eleven low-lying states of $^{22}$Na are 1.557, 0.661, and 1.077 MeV, respectively, for the INOY, CDB2K, and N$^3$LO interactions, in accordance with experimental data. The shell-model results using the USDB interaction presented in the same figure show good agreement with the experimental data up to the fourth excited state (5$^+_1$). However, some disagreement in terms of excitation energies and ordering of states is seen for the rest of the six states. 
	
	In the second panel of \autoref{spectra}, we have shown the low-energy spectra of $^{23}$Na. While the INOY interaction is able to reproduce the correct ordering of low-energy spectra up to the second excited state (7/2$^+$), the other two interactions, namely CDB2K and N$^3$LO, are able to reproduce only the g. s. (3/2$^+$) and first excited state (5/2$^+$) correctly.  For the case of $^{23}$Na, the rms deviations in energies of low-energy spectra comprising nine states are 0.903, 0.811, and 1.383 MeV, respectively, corresponding to the experimental data. On the other hand, the shell model results  presented in the same figure using USDB interaction show good agreement with the experimental data. 
	
	
	\begin{figure*}
		\centering
		\includegraphics[scale = 0.55]{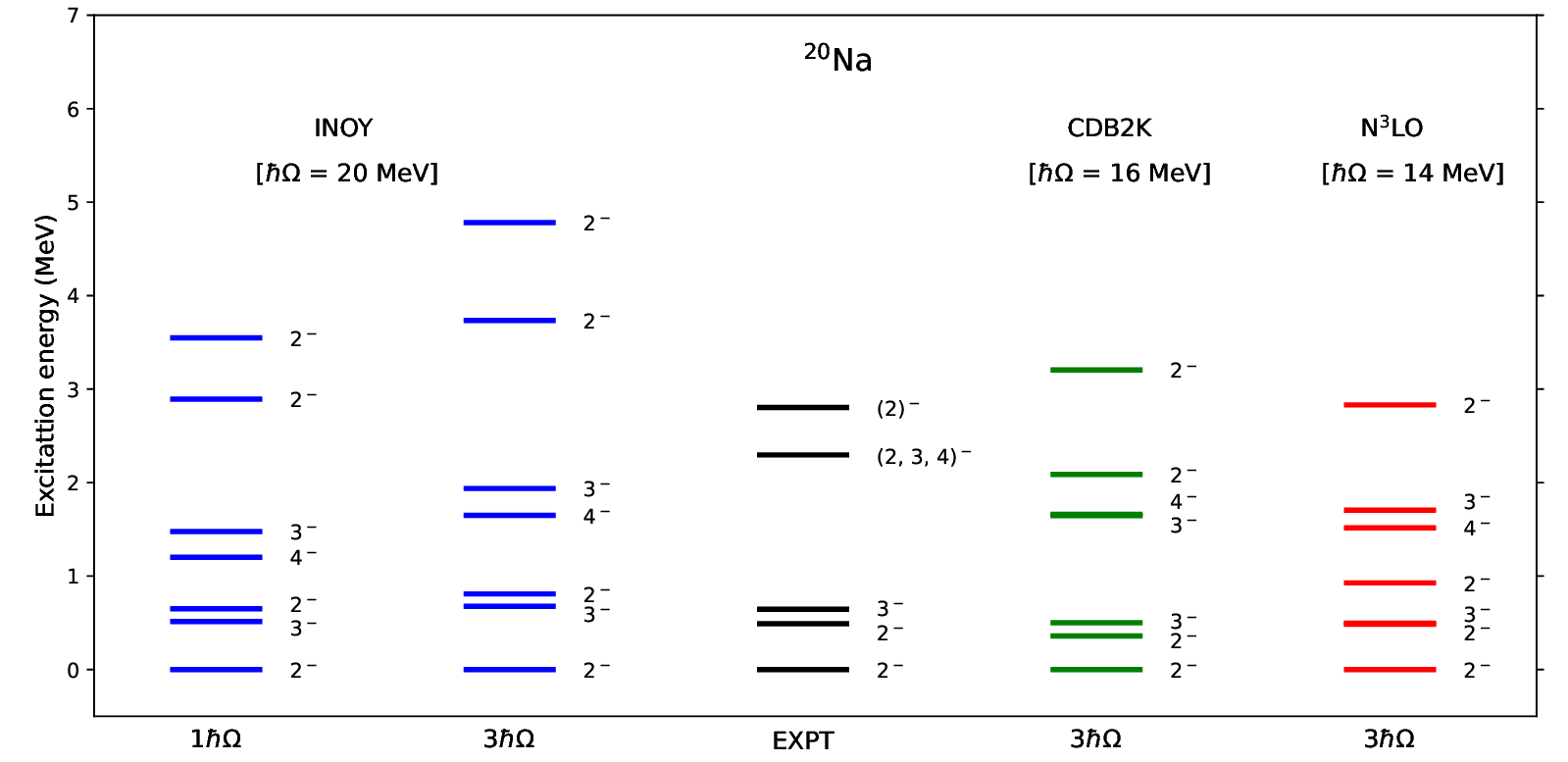}
		\includegraphics[scale = 0.55]{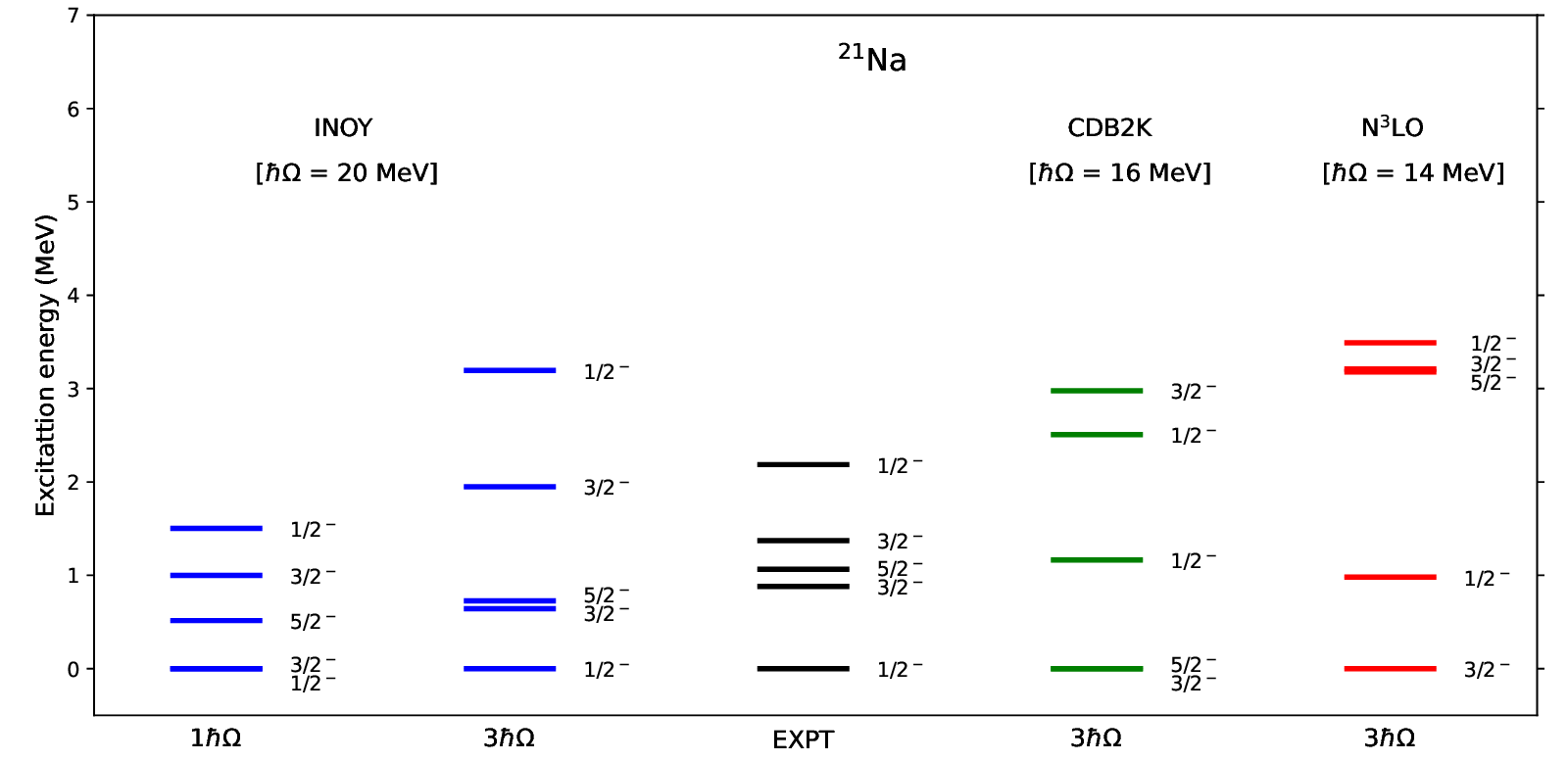}
		\includegraphics[scale = 0.55]{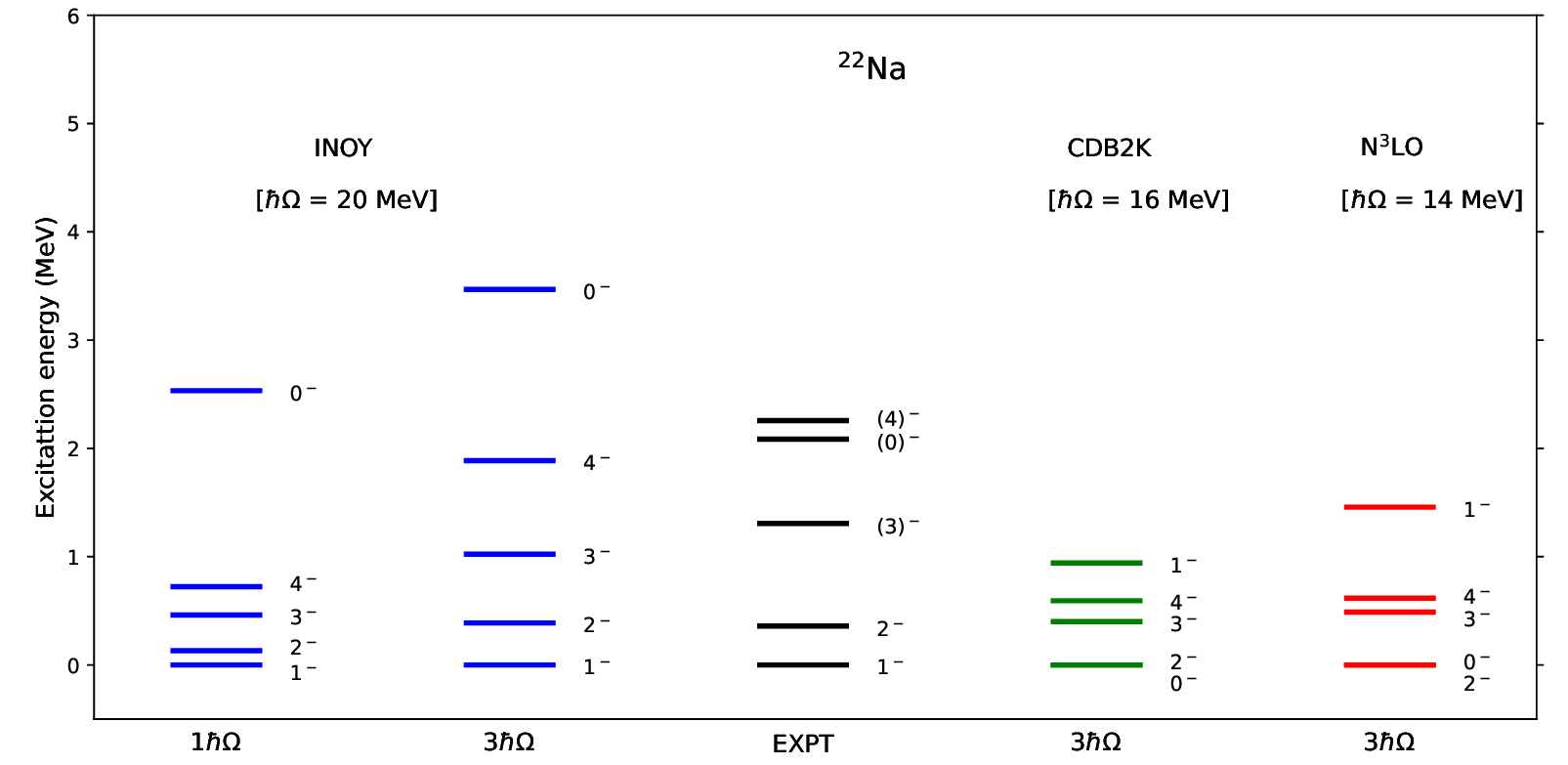}
		\caption{Low-energy un-natural parity states of $^{20, 21, 22}$Na.}
		\label{spectra3}
	\end{figure*}
	
	\begin{figure*}
		\centering
		\includegraphics[scale = 0.55]{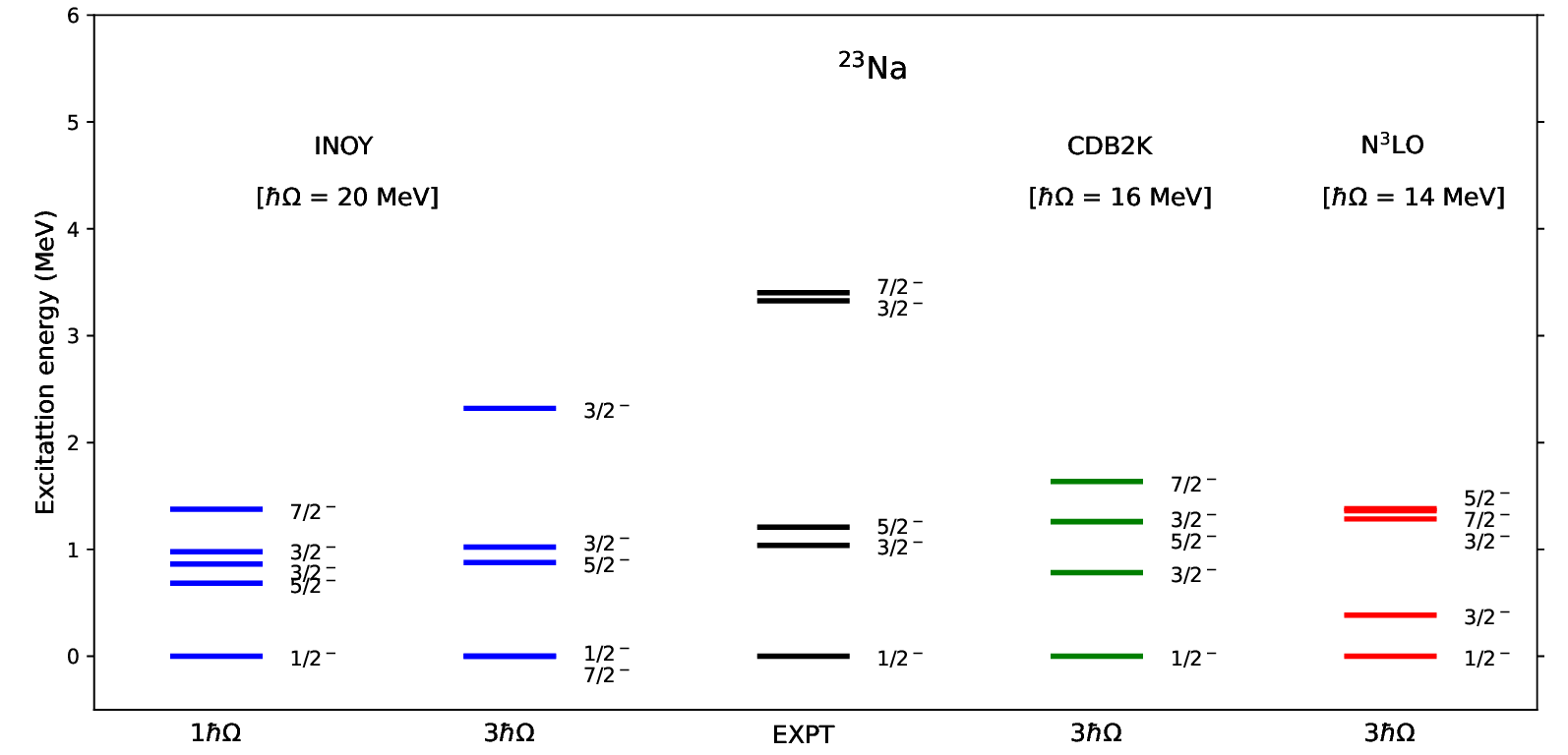}
		\caption{Low-energy un-natural parity states of $^{23}$Na.}
		\label{spectra4}
	\end{figure*}

	\subsection{Un-natural parity low-lying states of $^{20-23}$Na:}

	In this section, we are discussing the low-lying unnatural parity states of $^{20-23}$Na. The ground states of all four isotopes have positive parity, and the un-natural negative parity states are shown in \autoref{spectra3} and \autoref{spectra4}. {\color{black} Also, the rms deviations in energies of the un-natural parity low-energy spectra are calculated to test the quality of each interaction similar to the case of natural parity states. }The first panel of \autoref{spectra3} shows the negative parity states of $^{20}$Na. As can be seen from the figure, all three interactions are able to reproduce 2$^-$ as the correct lowest negative parity state. While CDB2K and N$^3$LO are able to reproduce the correct sequence of states up to 3$^-_1$, the ordering of 2$^-_2$ and 3$^-_1$ are seen to be reversed for INOY interaction. The 2$^-_4$ state obtained for N$^3$LO interaction is in good agreement with the un-confirmed (2)$^-_4$ state, so the 4.150 MeV state of $^{20}$Na negative parity spectra could be 2$^-$ as predicted from our NCSM calculation. {\color{black} For the case of $^{20}$Na, only three low-lying un-natural states are considered to calculate the rms deviation in energies as the spin-parity of the third excited state is not confirmed in the available data. The rms energy deviations are 0.226, 0.139, and 0.107 MeV, respectively, for INOY, CDB2K, and N$^3$LO interactions corresponding to the experimental data.}
	
	The second panel of \autoref{spectra3} shows the un-natural parity states of $^{21}$Na. From the figure, we can see that the sequence of states is correctly reproduced by INOY interaction. While the calculated excitation energies of 3/2$^-_1$ (5/2$^-_1$) states are less by 240 keV (337 keV) compared to the experimental states, they are higher by 576 keV (1008 keV) for 3/2$^-_2$ (1/2$^-_1$) states compared to the experimental data. The other two interactions fail to reproduce the correct ordering of low-energy spectra including the g.s. {\color{black} The rms energy deviations are 0.551, 0.958, and  1.437 MeV for the five low-lying un-natural parity states of $^{21}$Na for INOY, CDB2K, and N$^3$LO, respectively, compared to the experimental data.}
	
	The negative parity spectra of $^{22}$Na are shown in the third panel of \autoref{spectra3}. While the INOY interaction is able to reproduce the correct ordering up to 3$^-_1$ state, the other two interactions fail to reproduce the correct spectra including the g.s. The calculated 2$^-_1$ state with INOY is in good agreement with the experimental data.  {\color{black} For the case of $^{22}$Na, the energy deviations for INOY, CDB2K, and N$^3$LO are  0.653, 1.270, and 1.252 MeV, respectively, corresponding to the five low-energy un-natural parity states.}

	The low-energy negative parity states of $^{23}$Na are shown in \autoref{spectra4}. The CDB2K and N$^3$LO interactions are able to reproduce the g.s. correctly, while the calculated results with INOY show 7/2$^-_1$ as the lowest negative parity state. The calculated spectra using CDB2K match the experimental ordering of states. The excitation energy of the 5/2$^-_1$ state for CDB2K is only 50 keV higher than the experimental data; however, the calculated excitation energies of 3/2$^-_1$, 3/2$^-_2$ and 7/2$^-_1$ states for the same interaction are lower than the experimental data. {\color{black}    The energy deviations are higher than 1 MeV for all three interactions in the case of $^{23}$Na compared to the experimental data.}

 {\color{black} The rms deviations in the low-energy spectra of $^{20-23}$Na isotopes corresponding to the experimental data are shown in \autoref{tab:rms} for INOY, CDB2K, and N$^3$LO interactions. Based on the rms values, it can be concluded that the CDB2K interaction is better at reproducing the low-lying natural parity states of $^{20-23}$Na isotopes compared to the other two interactions. Nevertheless, a comprehensive inference regarding the unnatural parity states of these sodium isotopes cannot be made. Although the rms deviations exhibit lower values for INOY in the context of $^{21, 22}$Na when contrasted with the other two interactions, they demonstrate higher values for $^{20, 23}$Na in comparison to CDB2K and N$^3$LO.}

\begin{table}[ht]
		\centering
		\caption{ {\color{black} The rms deviations in the low-energy spectra (natural and un-natural) are shown for three $NN$ interactions considered in this work corresponding to the experimental data. All results are in MeV.}}
		\label{tab:rms}
		\begin{tabular}{cccccccc}
			\hline
			\hline 
            {\color{black}$^{20}$Na}  & \hspace{5mm} & {\color{black}INOY}  & \hspace{5mm} &  {\color{black}CDB2K}  & \hspace{5mm} &  {\color{black}N$^3$LO}\\
			\hline 
			{\color{black}Natural} &  & {\color{black}1.146}   &  & {\color{black}0.720} &  & {\color{black}0.941} \\
			{\color{black}Un-natural} &  & {\color{black}0.226} &  & {\color{black}0.139} &  & {\color{black}0.107}\\
			\hline
		{\color{black}$^{21}$Na}  &  &    &  &   &  &  \\
			\hline
			{\color{black}Natural} &  & {\color{black}1.645}  &  & {\color{black}0.375} &  & {\color{black}0.439} \\
			{\color{black}Un-natural} &  & {\color{black}0.551} &  & {\color{black}0.958} &  & {\color{black}1.437}\\
                \hline
		{\color{black}$^{22}$Na}  &  &    &  &   &  &  \\
			\hline
			{\color{black}Natural} &  & {\color{black}1.557}  &  & {\color{black}0.661} &  & {\color{black}1.077} \\
			{\color{black}Un-natural} &  & {\color{black}0.653} &  & {\color{black}1.270} &  & {\color{black}1.252}\\
                \hline
		{\color{black}$^{23}$Na}  &  &    &  &   &  &  \\
			\hline
			{\color{black}Natural} &  & {\color{black}0.955} &  & {\color{black}0.866} &  & {\color{black}1.277} \\
    			{\color{black}Un-natural} &  & {\color{black}1.782} &  & {\color{black}1.359} &  & {\color{black}1.445 }\\
			\hline \hline
		\end{tabular} 
	\end{table}
 
\begin{figure*}
		\centering
		\includegraphics[scale = 0.55]{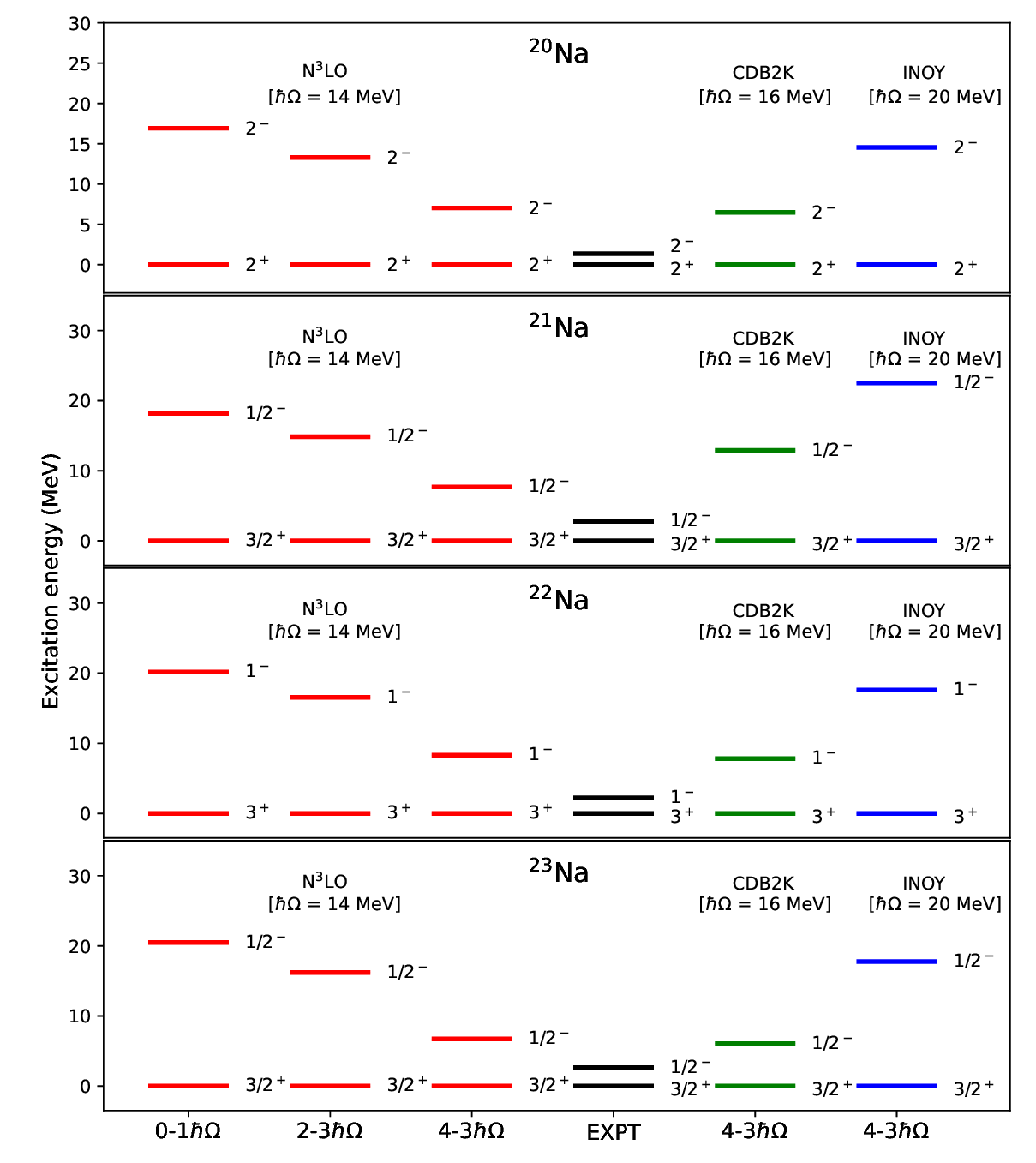}
		\caption{The excitation energies of the lowest un-natural parity states of $^{20-23}$Na isotopes are shown with respect to the g.s. For N$^3$LO, results are shown from model space from 0$\hbar \Omega$ to 4$\hbar\Omega$ and for the other two interactions, 3$\hbar\Omega$ and 4$\hbar\Omega$ results are shown. }
		\label{spectra5}
	\end{figure*}

	In \autoref{spectra5}, the lowest positive and negative parity states of $^{20-23}$Na isotopes are compared for the model space going from 0$\hbar\Omega$ to 4$\hbar\Omega$ for the case of N$^3$LO interactions. Calculated results corresponding to CDB2K and INOY are shown only for 3$\hbar\Omega$ and 4$\hbar\Omega$ model spaces. We observed that the excitation energies of the unnatural states improve by increasing the model space size for all four Na isotopes. Further improvements in excitation energies of un-natural parity states can be expected for even larger model space. Of the three interactions considered in this work, INOY shows {\color{black} higher} excitation energies of the corresponding negative parity states, even for the highest model spaces considered. The deviation with INOY interaction is very large corresponding to 0$\hbar\Omega$ to 1$\hbar\Omega$  and 2$\hbar\Omega$ to 3$\hbar\Omega$ model spaces. Because of this reason, we have shown N$^3$LO results corresponding to the above two model spaces in \autoref{spectra5}.

 States with un-natural parity exhibit lower binding energies than states with natural parity. The binding energies of the lowest un-natural parity state relative to the g.s. converge towards experimental data as the model space size increases, as depicted in \autoref{spectra5}. However, the convergence of un-natural parity spectra relative to the lowest natural parity state is slow with the basis size increment. In this work, the un-natural parity states are calculated using model spaces $N_{max}$ = 1 and 3, which are not large enough to get complete convergence for those negative-parity states. Furthermore, distinct interactions exhibit varied convergence characteristics for both natural and un-natural parity states. As a result, the computed outcomes depicted in \autoref{spectra3} and \autoref{spectra4} vary across three distinct interactions. Improved un-natural parity spectra and convergence of the separation between the lowest natural and un-natural parity states are anticipated when utilizing $N_{max}$ = 5 or greater model spaces. 
 Furthermore, employing a different  $\hbar\Omega$ value, other than that corresponding to the minima of the g.s. curve at $N_{max}$ = 4, may yield superior un-natural parity spectra for the four sodium isotopes specified.	
	
	\subsection{Mirror energy difference in the low-energy spectra of $A$ = 21 mirror pairs:}
	
	The isospin symmetry breaking can be studied by considering pairs of mirror nuclei where the number of protons and neutrons are interchanged. It is responsible for the difference in excited spectra of mirror pairs. Though the Coulomb interaction among the protons is the major source of isospin symmetry breaking, some contributions also come from the nuclear part. The mirror energy difference (MED) between the analogous states of mirror pairs are defined as:
	\begin{eqnarray}
	MED_J = E_J(T_z = -T) - E_J(T_z = +T).
	\end{eqnarray}
	Here, $E_J$ are the excitation energies of analogous states with angular momentum $J$ in a mirror pair with $T_z$ = $\pm T$. The MED provides an estimation of the isospin symmetry breaking of a particular interaction. 
	
\begin{figure*}
		\includegraphics[scale = 0.60]{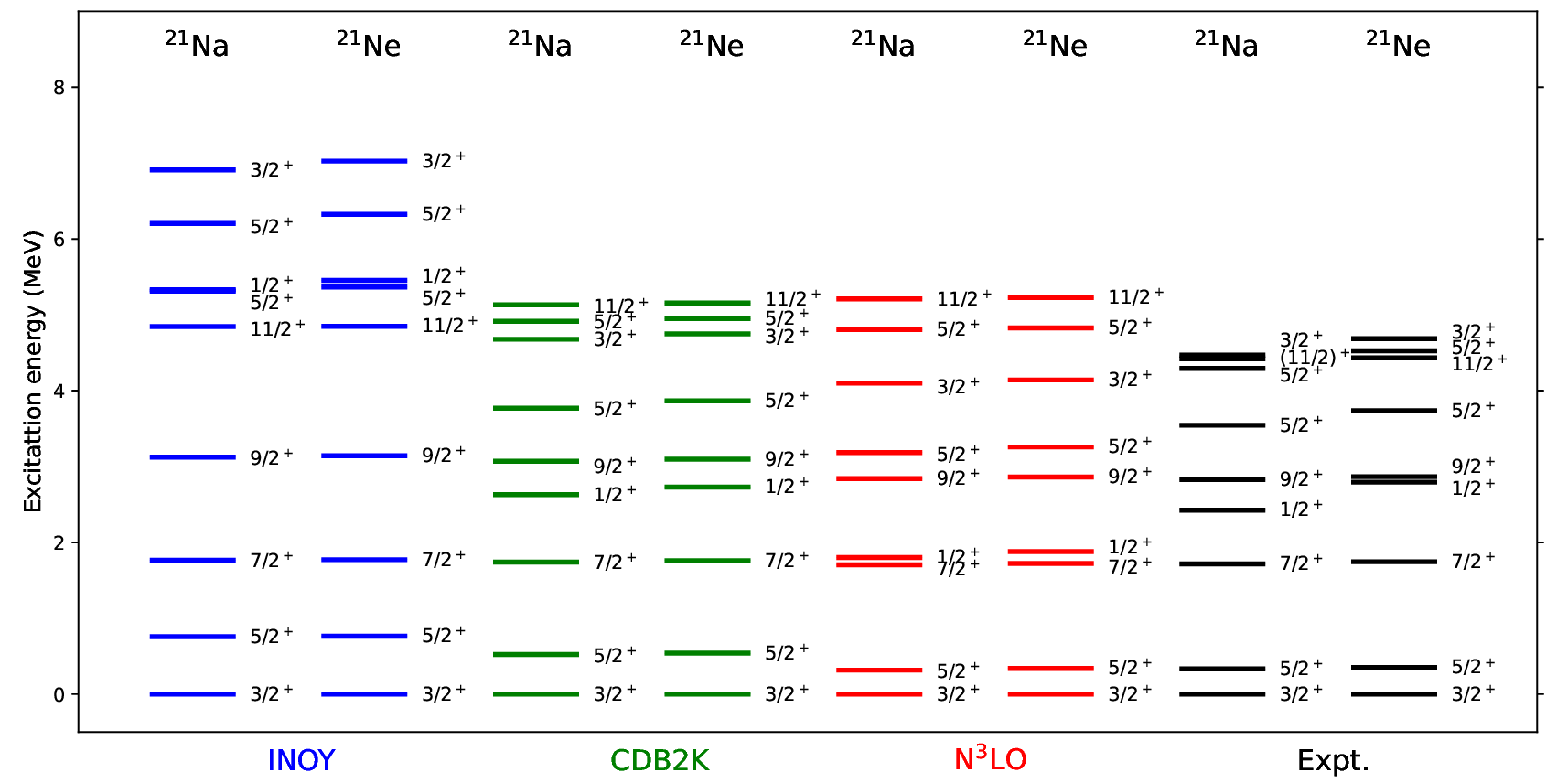}
		\caption{Comparison of the low-energy states of $^{21}$Na and $^{21}$Ne.}
		\label{21Na_MED}
	\end{figure*}

\begin{figure}
		\centering
		\includegraphics[scale = 0.50]{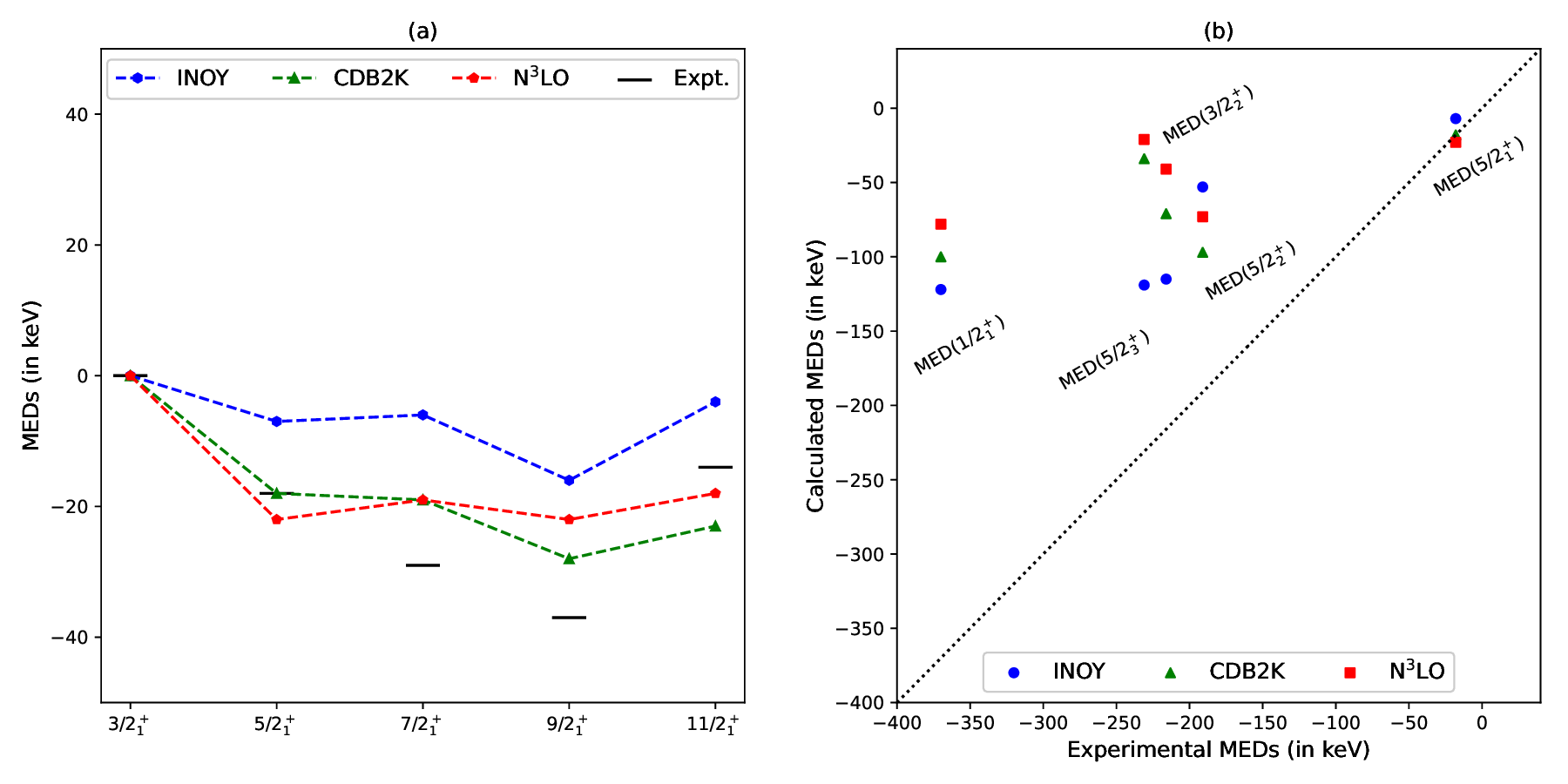}
		\caption{Mirror energy difference between the low-energy states of $^{21}$Na and $^{21}$Ne.}
		\label{21Na_MED2}
	\end{figure}

	In \autoref{21Na_MED}, we have shown a comparison between the low-energy states of $^{21}$Na and $^{21}$Ne which form a $|T_z|$ = 1/2 mirror doublet. The {\color{black} calculated} spectra of $^{21}$Ne are taken from Ref. \cite{chandan}. All results are for $N_{max}$ = 4 at the corresponding optimum frequencies of each interaction. {\color{black}The yrast band states: 3/2$^+_1$-5/2$^+_1$-7/2$^+_1$-9/2$^+_1$-11/2$^+_1$ are slightly different in energy (less than 50 keV) between the analogous states of the two mirror pairs as shown  in the left panel of \autoref{21Na_MED2}. From this figure, it can be seen that the difference between the calculated and the experimental MEDs corresponding to these yrast band states do not exceed 20 keV. However, some other low-energy states show large MEDs, and for those states, large deviations between the calculated and experimental MEDs can be seen. In the  right panel of \autoref{21Na_MED2}, we show the comparison between the experimental and calculated MEDs of some of the low-lying states, which show a large MED}. For comparison, we also include MED of 5/2$^+_1$, which is a yrast band state.  Experimentally, a small MED (18 keV) is seen for 5/2$^+_1$, while the 1/2$^+_1$ shows a large MED of 370 keV. The calculated MEDs for this state (1/2$^+_1$) are 122, 100, and 78 keV with INOY, CDB2K and N$^3$LO, respectively. The other three states, 3/2$^+_2$, 5/2$^+_2$, and 5/2$^+_3$ show MEDs $\sim$ 200 keV. The calculated MEDs with all three realistic interactions underestimate {\color{black}the corresponding experimental data}. It is observed that out of the three realistic interactions, INOY produces the highest MEDs between analogous states of the mirror pairs considered in this work. 
	

        \begin{figure}
		\includegraphics[scale=0.48]{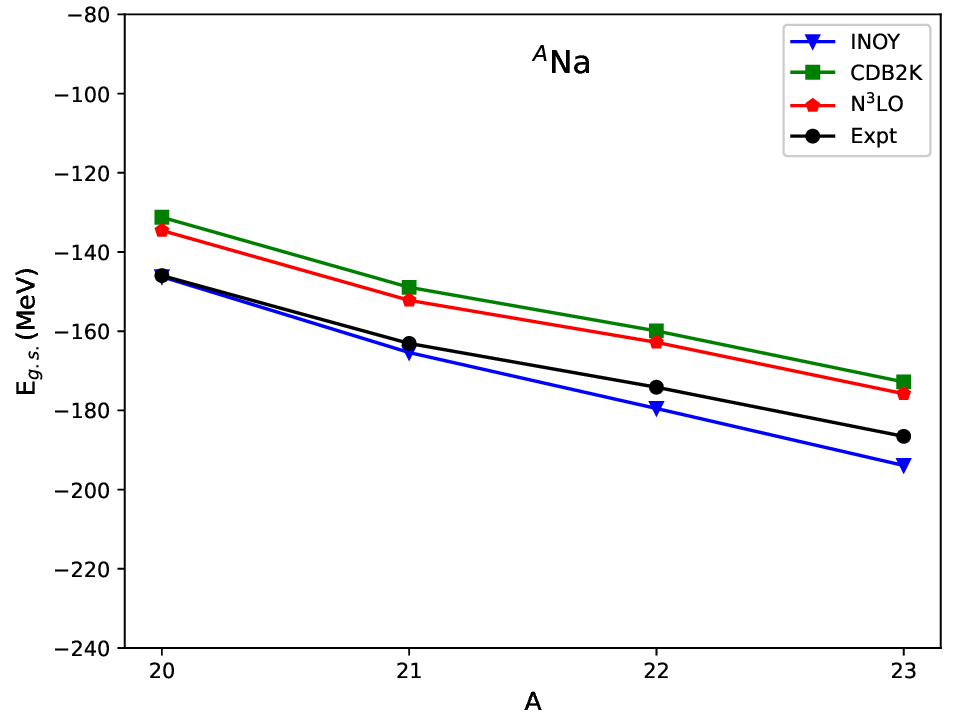}
		\caption{Comparison of the calculated  g.s. energies of $^{20-23}$Na isotopes with the experimental data.}
		\label{em1}
	\end{figure}

	\begin{figure*}
		\centering
		\includegraphics[scale = 0.50]{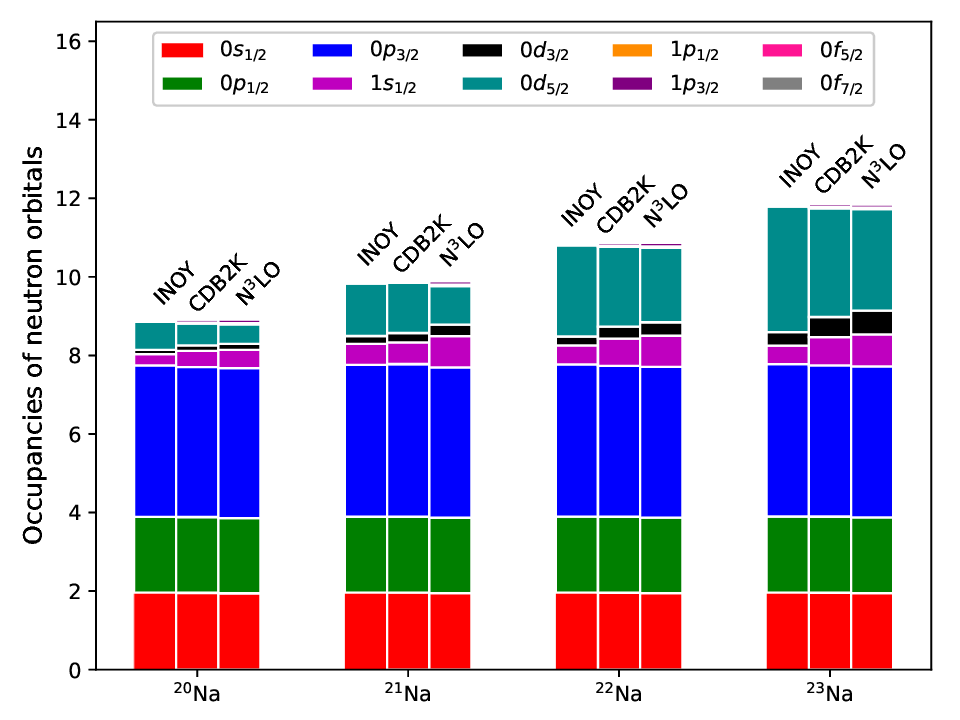}
		\includegraphics[scale = 0.50]{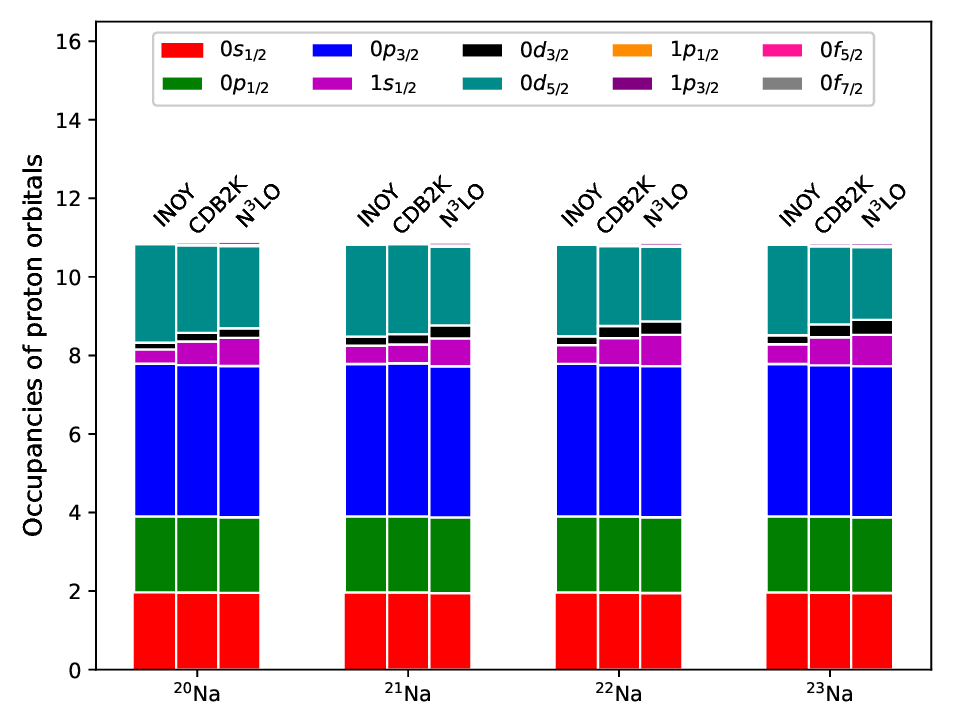}
		\caption{Neutron and proton occupancies of $^{20-23}$Na isotopes are shown for INOY, CDB2K and N$^3$LO interactions.}
		\label{gs_occupancy}
	\end{figure*}
	
	\subsection{Ground state energies and electromagnetic properties}
	
	In this section,  we discuss the binding energies of the  g.s. and the electromagnetic properties of all four sodium isotopes considered. All the results presented in \autoref{em} are with respect to the optimal frequencies of each interaction corresponding to the highest $N_{max}$ calculation. The experimental data included in the table are taken from Refs. \cite{NNDC,Qandmag}. {\color{black}Additionally, the shell-model results using the USDB \cite{usdb} interaction are also included in this table. The effective charges of $e_p = 1.36e$ and $e_n = 0.45e$ are considered for {\color{black} $E2$ transition strengths and quadrupole moments.}
	
	We first discuss the binding energies of the g.s of {\color{black} Na isotopes}. As shown in \autoref{em}, {\color{black}the g.s. binding energy of $^{20}$Na} is -145.960 MeV and the calculated result with INOY interaction overbind the {\color{black} g.s.} by only 281 keV. While the other two interactions significantly underbind the {\color{black} g.s.} of $^{20}$Na. In \autoref{em1}, we have plotted the {\color{black} g.s.} binding energies of sodium isotopes obtained from all three interactions considered in this work. From the figure, it is seen that the INOY interaction overbinds the {\color{black} g.s.} of all four {\color{black} Na isotopes}, while the other two interactions underbind corresponding {\color{black} g.s.} significantly. On average, while the INOY interaction overbinds the {\color{black} g.s.} by 5.140 MeV, the CDB2K and N$^3$LO underbind by 13 and 11 MeV, respectively.  
    The neutron and proton occupancies up to {\color{black}0$f_{7/2}$} orbital are shown in \autoref{gs_occupancy} for the {\color{black} g.s.} of $^{20-23}$Na isotopes corresponding to three different interactions. From the figure \autoref{gs_occupancy}, a significant difference is seen for the occupancies of $\nu (1s_{1/2})$ and $\pi (1s_{1/2})$ orbitals for INOY and N$^3$LO interactions. These differences in occupancies might have caused the differences in the binding energies of the {\color{black} g.s.} of $^{20-23}$Na isotopes for INOY and N$^3$LO {\color{black} interactions}. Lower occupancies of 1s$_{1/2}$ {\color{black} orbital} for INOY can be correlated to more binding of the {\color{black} g.s.} compared to CDB2K and N$^3$LO interactions.

To reproduce the g.s. binding energies, we need 3$N$ forces. However, the INOY interaction, due to its short-range non-local part, can reproduce the {\color{black} g.s.}  binding energies close to the experimental data without any 3$N$ interaction. The other two interactions, namely CDB2K and N$^3$LO, lack additional 3$N$ force, and the {\color{black} g.s.} binding energies for these two interactions can deviate up to 10\% of experimental binding energy as expected. 
From the natural parity spectra of $^{21}$Na (second panel of \autoref{spectra1}), the 5/2$^+_1$ state is pushed upward for the case of INOY compared to the other two interactions, suggesting an enhanced spin-orbit interaction strength. Similarly, in the case of $^{23}$Na (\autoref{spectra}), the spin-orbit interaction strength is almost the same for INOY and CDB2K, while it is less for the N$^3$LO interaction. 
Also, from the occupancy plots in \autoref{Pi_occupancy}, it can be seen that while $N_{max}$ = 0 model space contributes 61.7\% to the {\color{black} g.s.} of $^{23}$Na for INOY interaction, the same is only 55.0 and 50.6\% for CDB2K and N$^3$LO interaction, respectively. 
This suggests that the tensor force concerning the OLS renormalized N$^3$LO interaction is stronger than INOY and CDB2K. Due to the moderate strengths for both spin-orbit and tensor force, the CDB2K interaction can reproduce better (less rms deviation) natural parity states of $^{20-23}$Na isotopes compared to the other two interactions. 
	
	Apart from the binding energies of the {\color{black} g.s.}, \autoref{em} shows the calculated quadrupole and magnetic moments of the {\color{black} g.s.} obtained from all three interactions. In \autoref{em2} and \autoref{em3}, we have plotted the calculated quadrupole and magnetic moments of the {\color{black} g.s.} of sodium isotopes, respectively, and compared them with the experimental data. The overall trends of {\color{black} g.s.} moments are followed by all three interactions, though there are some deviations between the calculated and the experimental values. The electromagnetic transition strengths, more specifically calculated $B(E2)$ values, are significantly less than the experimental values, as can be seen from \autoref{em}. The prime reason for this is that $B(E2)$ is a long-range operator. While the OLS transformation renormalizes the short-range part of nuclear interaction and short-range operators, it weakly renormalizes the long-range operators like $B(E2)$. 
	Thus, better results for $B(E2)$ can be obtained once we perform calculations with higher $N_{max}$. For comparison, we also included the valence-space in-medium similarity renormalization group (VS-IMSRG) results of $B(E2; 5/2^+_1 \to 3/2^+_1$) transitions for $^{21}$Na and $^{23}$Na isotopes in \autoref{em} {\color{black}from Ref.} \cite{vs_imsrg2}. For these two cases, the $B(E2)$ strengths obtained for VS-IMSRG are better than the NCSM results for all three interactions considered in this work. However, the calculated values of $E2$ transition strengths from VS-IMSRG and NCSM are still far from the experimental data. Out of the three interactions considered in our work, N$^3$LO reproduces better results for $B(E2)$ transition strengths.  On the other hand, the $B(M1)$ transition strengths depend on the spin and isospin coordinates only for which converged results can be achieved easily unlike $B(E2)$ transitions.

	\begin{figure}
		\includegraphics[scale=0.48]{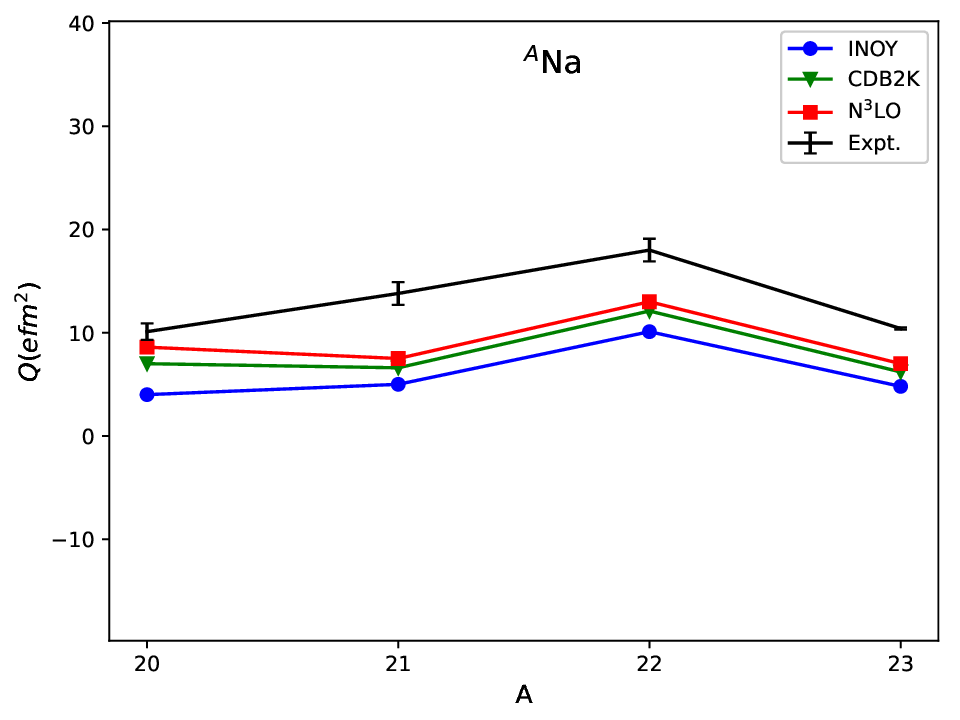}
		\caption{Comparison of the calculated {\color{black} g.s.} quadrupole moments of {\color{black} $^{20-23}$Na} isotopes with the experimental data.}
		\label{em2}
	\end{figure}
	
	\begin{figure}
		\includegraphics[scale=0.48]{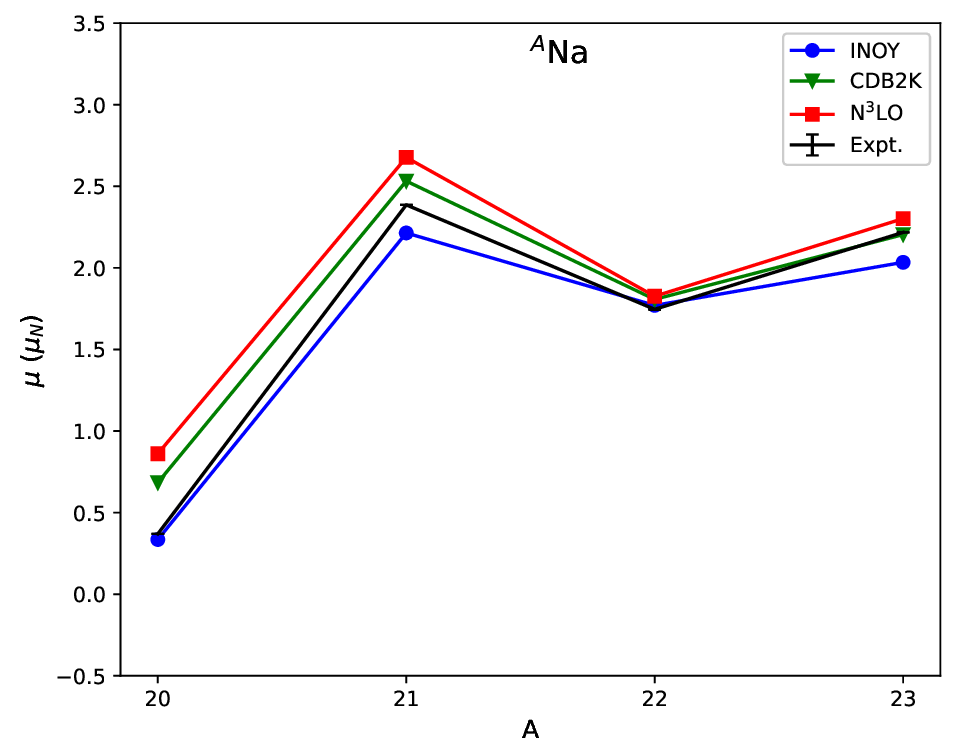}
		\caption{Comparison of the calculated {\color{black} g.s.} magnetic moments of {\color{black} $^{20-23}$Na} isotopes with the experimental data.}
		\label{em3}
	\end{figure}
	\begin{table*}
		\centering
		
		\caption{\label{em} The {\color{black} g.s.} energies and electromagnetic observables of $^{20-23}$Na corresponding to the highest $N_{\mbox{max}}$ at their optimal HO frequencies. The {\color{black} g.s.} energies, quadrupole moments, magnetic moments, $E2$ and $M1$ transitions are in MeV, {\color{black} $e$fm$^{2}$}, nuclear magneton ($\mu_{N}$), {\color{black} $e^{2}$fm$^{4}$ and $\mu_{N}^{2}$, respectively.} Experimental values are taken from Refs. \cite{NNDC,Qandmag}. The VS-IMSRG results for $B(E2)$ transitions are taken from Ref. \cite{vs_imsrg2}. {\color{black}The shell-model calculations were performed using the USDB \cite{usdb} interaction, considering effective charges of $e_p = 1.36e$ and $e_n = 0.45e$ for all calculations}.}
		\begin{tabular}{cMMMMMMM}
			\hline
			\hline \vspace{-2.8mm}\\
$^{20}$Na & EXPT & INOY & CDB2K & N$^3$LO & VS-IMSRG & {\color{black}USDB} \\
			\hline 
			$E_{g.s.}(2^{+})$ & -145.96 & {\color{black}-146.24} &  {\color{black}-131.22} &  {\color{black}-134.57} & - & -\\ 
            $Q(2^{+}$) & {\color{black}10.1(8)} &  {\color{black}4.0} & {\color{black}7.0} &  {\color{black}9.0} & - & {\color{black}8.0}\\
			$\mu$($2^{+}$) & 0.3694(2) & {\color{black}0.34} & {\color{black}0.68} & {\color{black}0.86} & - &{\color{black}0.45}\\
			$B(E2;4_{1}^{+}$ $\rightarrow$ $2_{1}^{+}$) & NA & {\color{black}6.33} & {\color{black}10.50} & {\color{black}13.08} & - & {\color{black}28.1}\\
			$B(M1;2_{1}^{+}$ $\rightarrow$ $3_{1}^{+}$) & NA & {\color{black}0.68} & {\color{black}0.12} & {\color{black}0.06} & - & {\color{black}0.53}\\
			\hline \vspace{-2.8mm}\\
$^{21}$Na & EXPT & INOY & CDB2K & N$^3$LO & VS-IMSRG & {\color{black}USDB}\\
			\hline \vspace{-2.8mm}\\
			$E_{g.s.}(3/2^{+})$ & -163.09 & -165.39 & -148.92  & {\color{black}-152.20} & - & -\\
            Q($3/2^{+}$) & {\color{black}13.8(11)} & {\color{black}5.0} & {\color{black}7.0} & {\color{black}7.0} & -&{\color{black}11.0}\\
			$\mu$($3/2^{+}$) & 2.38610(4) & {\color{black}2.21} & {\color{black} 2.53} & {\color{black}2.68} & - &{\color{black}2.49}\\
			$B(E2;5/2_{1}^{+}$ $\rightarrow$ $3/2_{1}^{+}$) & 134(10) & {\color{black}19.58} 
			& {\color{black}27.96} & {\color{black}33.12} & 56.1 & {\color{black}90.0}\\
			$B(E2;7/2_{1}^{+}$ $\rightarrow$ $5/2_{1}^{+}$) & 55(27) & {\color{black}12.53} & 
			{\color{black}22.93} & {\color{black} 30.27} & - & {\color{black}59.2}\\
			$B(E2;7/2_{1}^{+}$ $\rightarrow$ $3/2_{1}^{+}$) & 72(27) & {\color{black}7.90} & {\color{black}12.80} & {\color{black}16.68} & - & {\color{black}37.5}\\
			$B(M1;5/2_{1}^{+}$ $\rightarrow$ $3/2_{1}^{+}$) & 0.1513(18) & {\color{black}0.24} & {\color{black}0.19} & {\color{black}0.22} & - & {\color{black}0.21}\\
			$B(M1;7/2_{1}^{+}$ $\rightarrow$ $5/2_{1}^{+}$) & 0.36(9) & {\color{black}0.32} & {\color{black}0.20} & {\color{black}0.13} & - & {\color{black}0.39}\\
			\hline \vspace{-2.8mm}\\
$^{22}$Na & EXPT & INOY & CDB2K & N$^3$LO & VS-IMSRG & {\color{black}USDB}\\
			\hline \vspace{-2.8mm}\\
			$E_{g.s.} (3^{+})$ & -174.15 & {\color{black}-179.51} & {\color{black}-159.93}  & {\color{black}-162.82} & - & -\\
            $Q(3^{+})$ & {\color{black}18.0(11)} & {\color{black}10.0} & {\color{black}12.0} & {\color{black}13.0} & - & {\color{black}23.0}\\
			$\mu$($3^{+}$) & 1.746(3) &  1.77 & 1.81 & 1.83 & -& {\color{black}1.79}\\
			$B(E2;1_{1}^{+}$ $\rightarrow$ $3_{1}^{+}$) & 0.03457(29) &  $< $ 0.01 & 0.22 & 0.60 & - & $< $ {\color{black}0.01}\\
			$B(E2;5_{1}^{+}$ $\rightarrow$ $3_{1}^{+}$) & 19.0(15) & 3.87 & 6.12 & 7.15 & - & {\color{black}21.2}\\
			$B(M1;0_{1}^{+}$ $\rightarrow$ $1_{1}^{+}$) & 4.96(18) & 4.37 & 7.90 & 12.60 & - & {\color{black}6.05}\\
			\hline \vspace{-2.8mm}\\
$^{23}$Na  & EXPT & INOY & CDB2K & N$^3$LO & VS-IMSRG & {\color{black}USDB}\\
			\hline \vspace{-2.8mm}\\
			$E_{g.s.} (3/2^{+})$ & -186.55  & {\color{black}-193.90} &  {\color{black}-172.82}  & {\color{black}-175.79} & -& -\\
            $Q(3/2^{+})$ & {\color{black}10.4(1)} & {\color{black} 5.0} & {\color{black} 6.0} & {\color{black} 7.0} & -  & {\color{black} 11.0}\\
			$\mu$($3/2^{+}$) & 2.21750(3) & {\color{black}2.03} & {\color{black}2.20} & {\color{black}2.30} & - & {\color{black}2.10}\\
			$B(E2;5/2_{1}^{+}$ $\rightarrow$ $3/2_{1}^{+}$) & 124(23) & 
			{\color{black}18.38} & {\color{black}26.38} & {\color{black}30.53} & 56.9 & {\color{black}109.0}\\
			$B(E2;7/2_{1}^{+}$ $\rightarrow$ $3/2_{1}^{+}$) & 47.4(58) & {\color{black}6.95} & {\color{black}11.55} & {\color{black}14.67} & -  & {\color{black}38.2}\\
			$B(M1;5/2_{1}^{+}$ $\rightarrow$ $3/2_{1}^{+}$) & 0.403(25) & {\color{black}0.35} & {\color{black}0.30} &  {\color{black}0.31} & - & {\color{black}0.38}\\
			\hline
			\hline
		\end{tabular}
	\end{table*}

	\subsection{Point-proton radius and neutron skin}
	Apart from the low-energy spectra and electromagnetic properties, we also investigated the point-proton radii of sodium isotopes. The operator defining point-proton radius is a long-range operator similar to the $B(E2)$ operator, and both operators are sensitive to the long-range part of the nuclear wavefunctions. So, to obtain converged results for these observables, NCSM calculations with large $N_{max}$ model space are required. Starting with the experimental charge radius, the point-proton radii can be extracted by using the following formula:
	\begin{equation}
	\langle r^2 \rangle_{p} = \langle r^2 \rangle_{c} - \langle R_p^2 \rangle - \frac{N}{Z} \langle R_n^2 \rangle - \frac{3}{4 m_p^2}.\\
	\end{equation}
	
	Here, $\langle R_p^2 \rangle$ and $\langle R_n^2 \rangle$ are, respectively, the squared charge radius of proton and neutron. The 3/(4$m_p^2$) is the Darwin-Foldy term related to relativistic correction in natural units. The values are taken to be $\langle R_p^2 \rangle^{1/2}$ = 0.8783(86) fm, $\langle R_n^2 \rangle$ = -0.1149(27) fm$^2$ \cite{charge_radii} and 3/(4$m_p^2$) = 0.033 fm$^2$ \cite{C.Forseen}. The squared point-proton radius relative to the center of mass of all nucleons is defined as
	\begin{equation}
	r_p^2  = \frac{1}{Z} \sum_{i=1}^{Z} \lvert \vec{r_i} - \vec{R}_{CM}\rvert^2.
	\end{equation}
	The operator for point-proton radius, as shown in the above equation, is a two-body operator. It is reduced to a more suitable form having one-body and two-body operators to evaluate two-body matrix elements for this operator. The expectation value of the $r_p$ operator is then calculated similarly to the {\color{black} g.s.} energy calculation. A similar method is applied to calculate the point-neutron radii ($r_n$) of atomic nuclei; however, experimental measurement of $r_n$ is challenging, unlike $r_p$ measurement. 
	
 In order to obtain converged results for the radius operator, large $N_{max}$ {\color{black} calculations} are needed. However, in this work, we use model space only up to $N_{max}$ = 4. So, without going to other theoretically motivated extrapolation techniques, the crossover prescription \cite{rp_1, rp_2} is used for our case to obtain converged $r_p$ and $r_n$. 
The radius corresponding to that point can be treated as a converged radius before achieving the true convergence with a large model space. In \autoref{rp}, the calculated $r_p$ and $r_n$ of $^{21, 22}$Na are shown as a function of $\hbar \Omega$ for $N_{max}$ = 2 and 4 model spaces. These figures show that at a lower HO frequency, the calculated $r_p$ and $r_n$ decrease with increasing $N_{max}$, while at a high HO frequency, they increase with increasing $N_{max}$. So, a region of calculated $r_{p}$ and $r_{n}$ around the crossing points is observed, which is independent of $N_{max}$. 
In this work, we take the intersection points of two $r_p$ and $r_n$ curves corresponding to $N_{max}$ = 2 and 4 as the converged radii. For example, the first panel of \autoref{rp} shows that the $r_p$ curves of $^{21}$Na for INOY interaction crosses each other almost at 2.10 fm. Similarly, the crossing point for $r_n$ curves is at 2.68 fm. So, the point-proton and point-neutron radii of $^{21}$Na corresponding to INOY interaction are 2.10 and 2.68 fm, respectively. Similarly, the converged $r_p$ and $r_n$ are calculated for all four sodium isotopes corresponding to the three realistic interactions considered. The calculated $r_p$ from the NCSM method are reported in \autoref{tab-rp} and compared with the experimental data which are taken from Ref. \cite{charge_radii}. From \autoref{rp_final}, we see that the converged $ r_p$ are underpredicted by NCSM results with all three interactions. While the N$^3$LO interaction shows the correct trend of experimental $r_p$, the INOY and CDB2K show odd-odd sodium isotopes $^{20, 22}$Na to have large $r_p$ compared to odd (Z) - even (N)  sodium isotopes $^{21, 23}$Na. Out of the three realistic interactions, the N$^3$LO reproduces slightly better results compared to the other two {\color{black} interactions}. This is due to large occupancies of the $\pi (1s_{1/2}$) orbital for N$^3$LO interaction compared to INOY and CDB2K interactions as can be seen from the {\color{black} right panel of} figure of \autoref{gs_occupancy}.

	\begin{figure*}
		\centering
		\includegraphics[scale = 0.50]{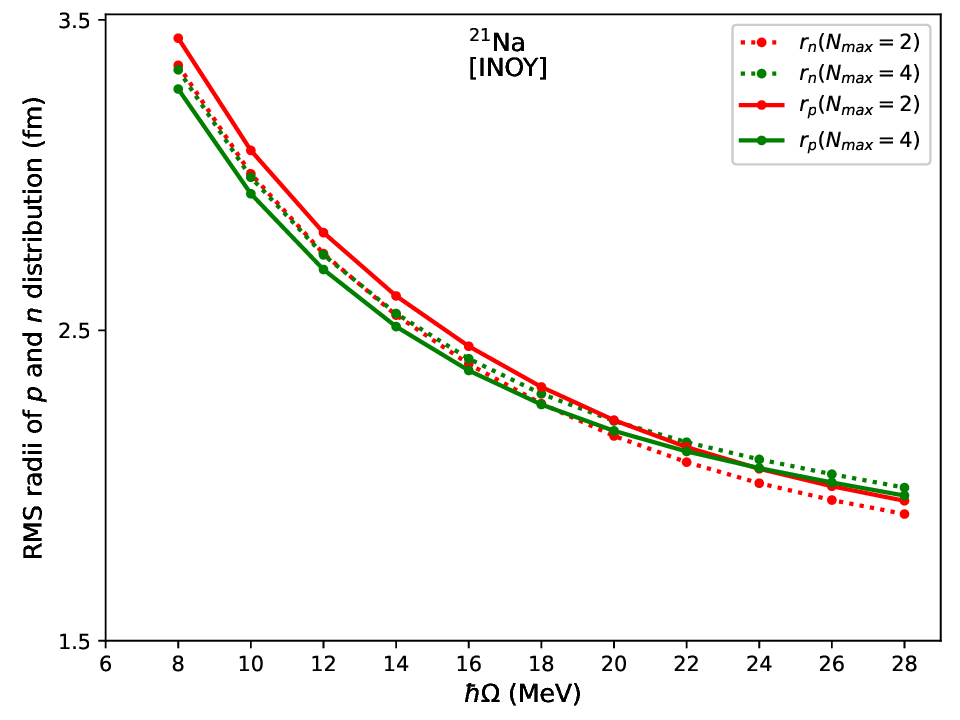}
		\includegraphics[scale = 0.50]{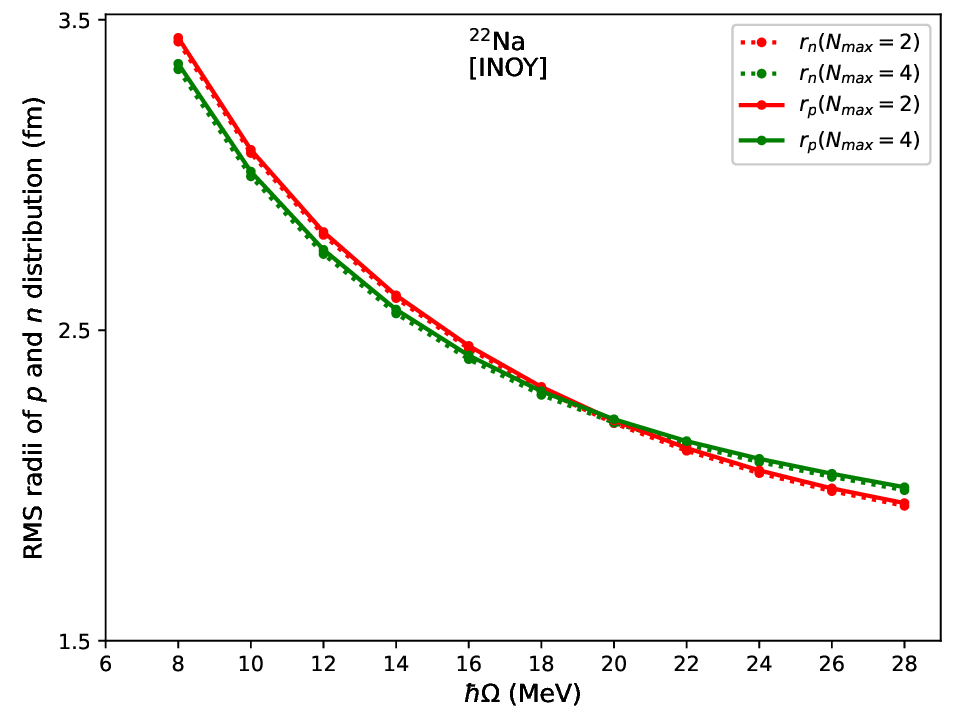}
		\includegraphics[scale = 0.50]{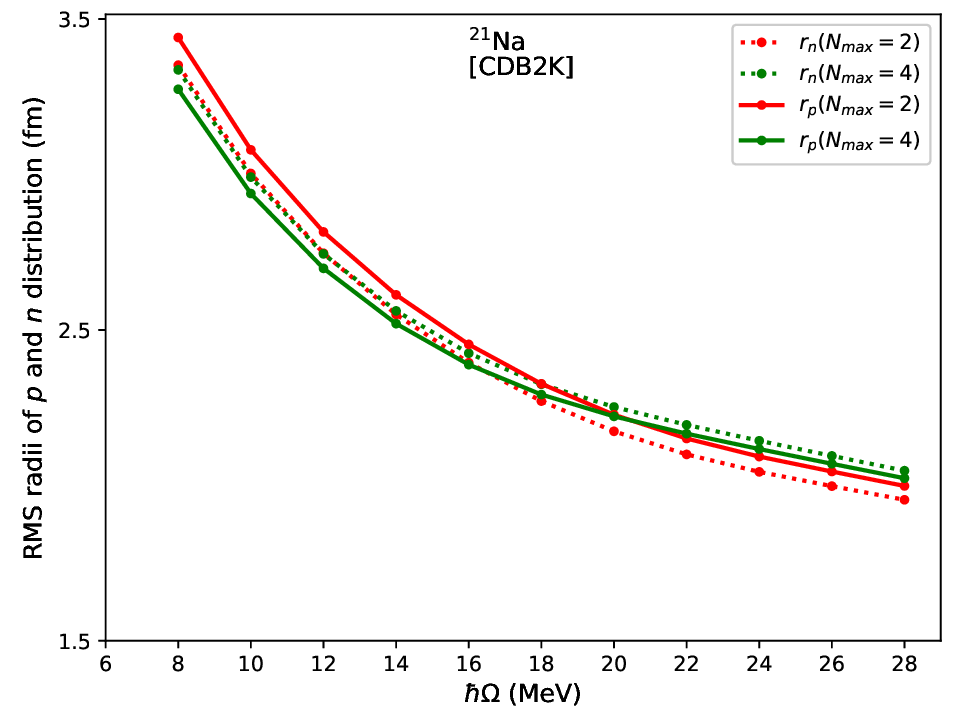}
		\includegraphics[scale = 0.50]{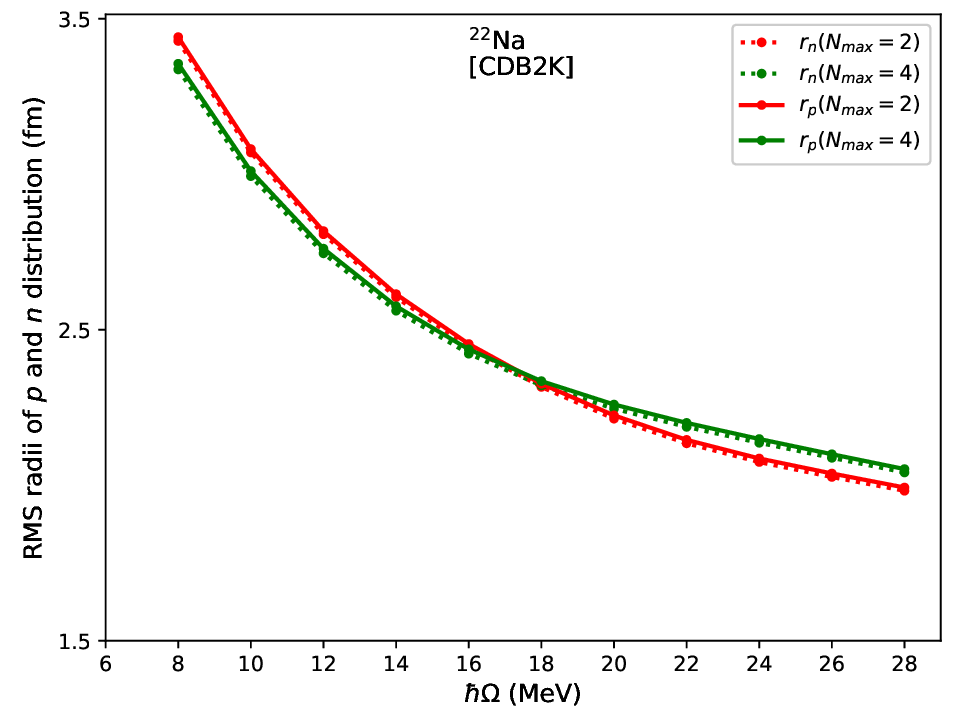}
		\includegraphics[scale = 0.50]{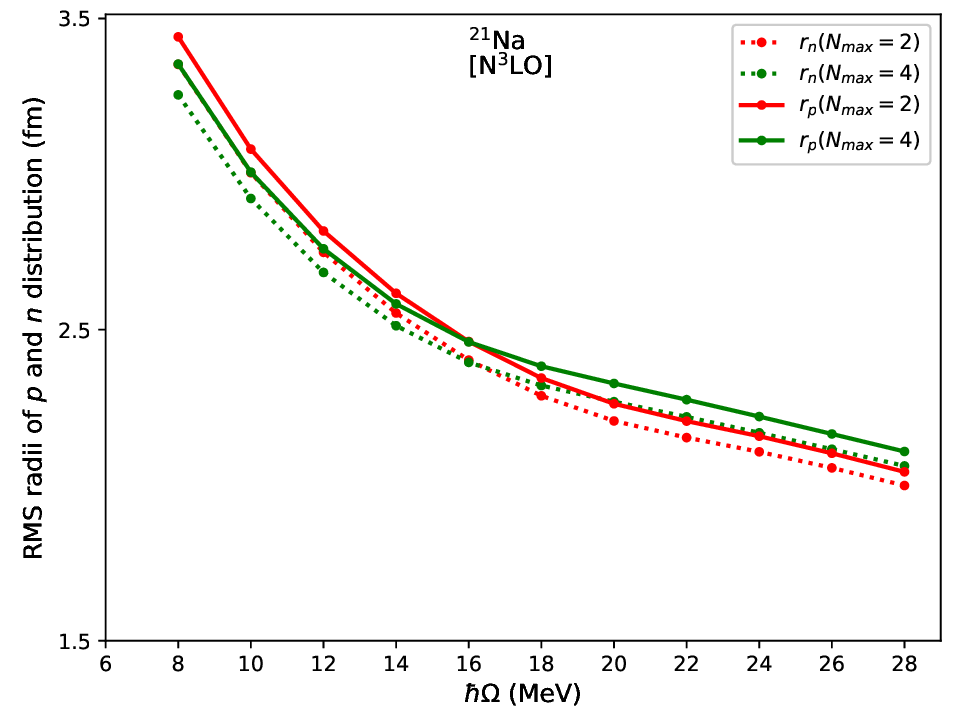}
		\includegraphics[scale = 0.50]{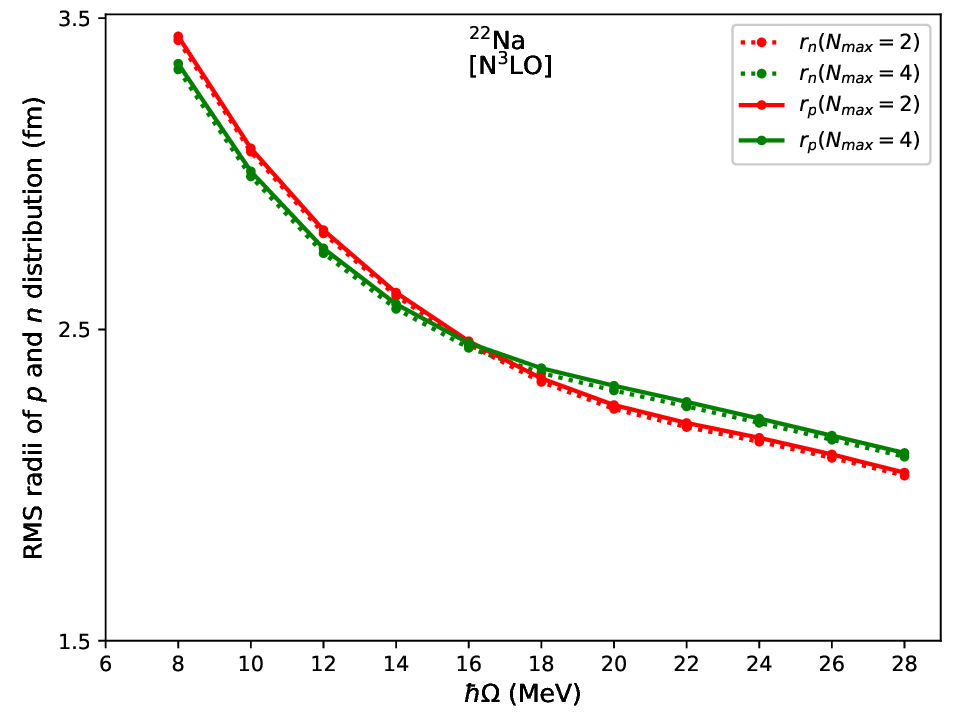}
		\caption{Root mean square radii of point-proton (solid line) and point-neutron (dotted line) distribution as a function of HO frequency for different $N_{max}$.}
		\label{rp}
	\end{figure*}
	
	\begin{table}[ht]
		\centering
		\caption{\label{tab-rp} The point-proton radii (r$_p$) of $^{20-23}$Na isotopes calculated with INOY, CDB2K and N$^3$LO are shown. The experimental data are taken from Ref. \cite{charge_radii}. {\color{black} All results are given in fm.}}
		\begin{adjustbox}{width=0.5\textwidth}	
			\begin{tabular}{cMMMM}
				\hline
				\hline 
				r$_p$ & Expt. & INOY & CDB2K & N$^3$LO\\
				\hline \vspace*{.5mm}
				$^{20}$Na & 2.8494(445) & 2.30 & 2.43 & 2.48\\
				$^{21}$Na & 2.8951(308) & 2.10 & 2.21 & 2.46\\
				$^{22}$Na & 2.8674(20) & 2.44 & 2.37 & 2.45\\
				$^{23}$Na & 2.8779(104) & 2.22 & 2.36 & 2.41\\
				\hline
				\hline	
			\end{tabular}
		\end{adjustbox}
	\end{table}
	
	From the knowledge of both $r_p$ and $r_n$, the neutron skin thickness, $r_{np}$ can be calculated using $r_{np}$ = $r_n$ - $r_p$. In \autoref{rp_final2}, the neutron skin thicknesses of the sodium isotope ground states are shown. The experimental data for the figure is taken from \cite{Na_skin}. From the figure \autoref{rp_final2}, we see that the calculated $r_{np}$ of $^{20,23}$Na {\color{black} g.s.} are within the experimental error for all three interactions. While $r_{np}$ of $^{22}$Na is in good agreement with the experimental data for INOY, for the other two interactions, they are slightly away from the experimental data. A significant mismatch is observed for $^{21}$Na $r_{np}$ corresponding to INOY and CDB2K interactions. However, the N$^3$LO result for the same is close to the experimental value. 
	
	\begin{figure}
		\centering
		\includegraphics[scale = 0.50]{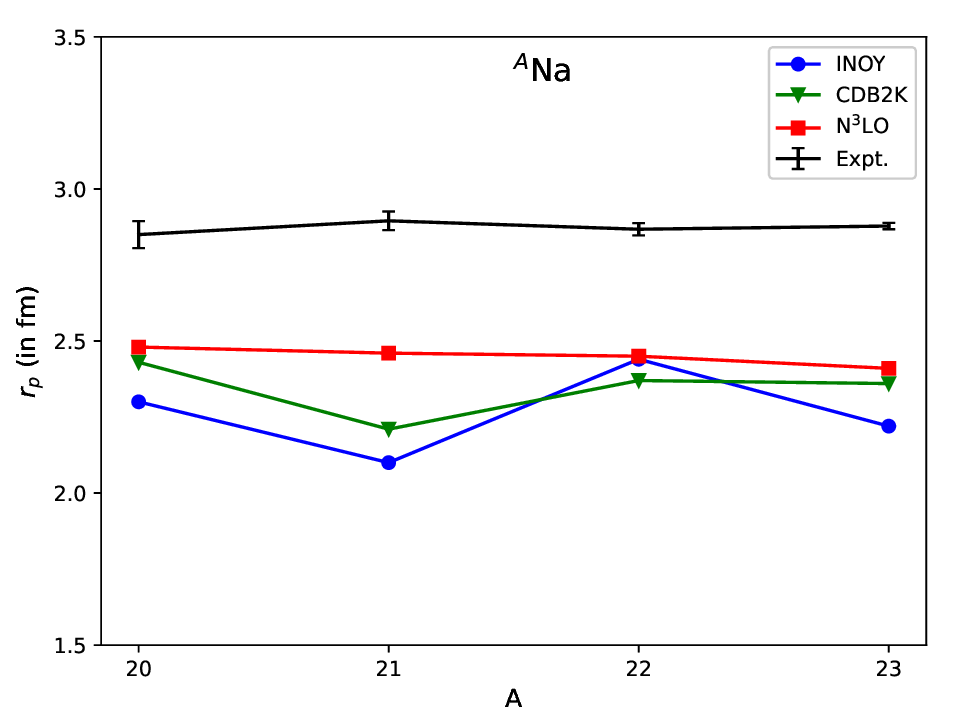}
		\caption{Comparison of the calculated {\color{black} g.s.} point-proton radii ($r_p$) of Na isotopes with the experimental data.}
		\label{rp_final}
	\end{figure}
	
	\begin{figure}
		\centering
		\includegraphics[scale = 0.50]{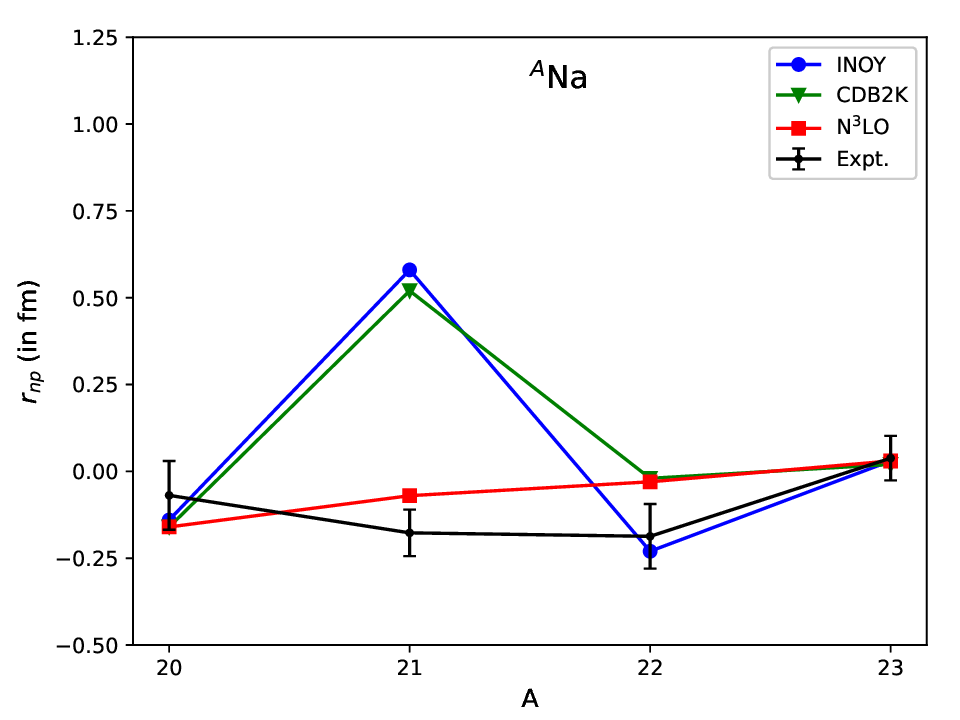}
		\caption{Comparison of the calculated {\color{black} g.s.} neutron skin thickness ($r_{np}$) of Na-isotopes with the experimental data.}
		\label{rp_final2}
	\end{figure}

 {\color{black} From the above discussions on different nuclear observables within the NCSM formalism utilizing three different classes of realistic interactions, it is worth mentioning that each interaction has some merits over the other interactions for certain observables. While the INOY interaction due to its non-local part, is able to reproduce close to the experimental {\color{black}g.s.} binding energies, the one boson exchange interaction CDB2K is able to reproduce better results for the natural parity states. On the other hand, the chiral interaction N$^3$LO is better at reproducing the $E2$ transition strengths and point-proton ($r_p$) radii of the four Na isotopes. These general observations showcase the complementary advantages offered by each interaction across different nuclear observables.} 
 
 {\color{black} Nuclear observables such as radii and transition strengths are highly sensitive to the wavefunctions. As each interaction provides a unique wave function for a particular state of a nucleus, it would provide different values for these observables corresponding to different interactions. In \autoref{gs_occupancy}, the occupancies of different orbitals are shown for the {\color{black}g.s.} of four {\color{black} Na isotopes} corresponding to INOY, CDB2K, and N$^3$LO, respectively. {\color{black} \autoref{gs_occupancy} shows the difference in the occupancies of harmonic oscillator orbitals corresponding to different interactions.}
	
	\section{Conclusions}
	\label{sect 5}
	In this work, we have investigated the low-lying nuclear structure properties of  {\color{black} $^{20-23}$Na} {\color{black} isotopes} within the \textit{ab initio} NCSM formalism using three realistic interactions, namely INOY, CDB2K, and chiral N$^3$LO. We studied the low-energy spectra of {\color{black} $^{20-23}$Na} , including both natural and unnatural parity states, electromagnetic properties, point-proton radii, and neutron skin thicknesses. We observed a good agreement of the g.s. binding energy of $^{20}$Na for INOY interaction with the experimental data. However, for other {\color{black} $^{21-23}$Na}, INOY interaction overbinds the corresponding g.s. The CDB2K and N$^3$LO underbind the g.s. of all four sodium isotopes. The INOY interaction results for the g.s. energies are better than  CDB2K  and  N$^3$LO interactions, this is because three-body force effects are absorbed in the nonlocal part of the INOY interaction.
	
	Among the electromagnetic properties, the quadrupole and magnetic moments of the g.s. follow the same trend as in the experimental data. The $B(M1)$ transition being independent of spatial coordinates, converged results close to experimental data can be achieved at a smaller basis space. However, the situation is different for the $B(E2)$ transition that depends on the long-range part of  wavefunctions. Among the three realistic interactions, the N$^3$LO reproduces better results for $E2$ transition strength for all four isotopes. However, only one-third of the experimental transition strengths are obtained for N$^3$LO interactions. Comparing with $B(E2)$ results of another \textit{ab initio} method, the $B(E2; 5/2_1^+ \to 3/2_1^+)$ for $^{21}$Na and $^{23}$Na are 56.1 and 56.9 $e^2 fm^4$ for VS-IMSRG, the NCSM calculation with N$^3$LO provide 33.12 and 30.56 $e^2 fm^4$. Compared to the experimental results, the $ab $ $initio$ results including VS-IMSRG are significantly less.
	
	The point-proton radii ($r_p$) is also a long-range observable, just like the $B(E2)$. In order to obtain the converged $r_p$, we employed the ``crossing-point'' method, and the converged $r_p$ for three different interactions are compared to the available experimental data. We observed that the g.s. $r_p$ obtained from NCSM calculation are less than the experimental data. However, among the three interactions, N$^3$LO reproduces slightly better results compared to the other two interactions and follows the same experimental trend.
	Similarly, we also calculated the converged root mean square radius of the neutron distribution ($r_n$) and the neutron skin thickness ($r_{np}$) of the {\color{black} g.s.} of  $^{21-23}$Na isotopes. Except for the case of $^{21}$Na, the calculated $r_{np}$ are in good agreement with the experimental data. 
	
	\section*{ACKNOWLEDGMENTS}
	We acknowledge financial support from SERB (India), CRG/2019/000556. We would like to thank Prof. Petr Navrátil for providing us his NN effective interaction code and Prof. Christian Forss\'en for making available the pAntoine.  We would also like to thank Prof. Ruprecht Machleidt for valuable comments on this article.

	

\end{document}